\documentclass[12pt]{iopart}
\usepackage{iopams}

\usepackage{amssymb}
\usepackage{graphicx}
\usepackage{epsfig}
\usepackage{color}
\usepackage{url}
\usepackage{times}
\usepackage{bm}
\usepackage{mathrsfs}
\usepackage[utf8]{inputenc}
\usepackage{hyperref}
\usepackage{enumerate}
\usepackage{amsthm}
\usepackage{verbatim}
\usepackage{cite}
\usepackage{bbm}
\usepackage{stmaryrd}
\usepackage{slashed}

\usepackage{caption}

\newcommand{\beq}{\begin{equation}}
\newcommand{\eeq}{\end{equation}}
\newcommand{\bea}{\begin{eqnarray}}
\newcommand{\eea}{\end{eqnarray}}
\newcommand{\bit}{\begin{itemize}}
\newcommand{\eit}{\end{itemize}}
\newcommand{\ben}{\begin{enumerate}}
\newcommand{\een}{\end{enumerate}}
\newcommand{\nn}{\nonumber}
\newcommand{\dfrac}[2]{{\displaystyle\frac{#1}{#2}}}
\newcommand{\eqref}[1]{(\ref{#1})}

\def\scri{\mathscr{I}}

\renewcommand{\d}{\,{\rm d}}

\newcommand{\QMUL}{\address{School of Mathematical Sciences, Queen Mary, University of
  London, \\ Mile End Road, London E1 4NS, United Kingdom}}
          
\theoremstyle{plain}

\newcounter{mnotecount}

\newcommand{\mnotex}[1]
{\protect{\stepcounter{mnotecount}}$^{\mbox{\footnotesize $\bullet$\themnotecount}}$ 
\marginpar{
\raggedright\tiny\em
$\!\!\!\!\!\!\,\bullet$\themnotecount: #1} }

\begin{document}
\title{Hyperboloidal framework for the Kerr spacetime
}
\author{Rodrigo Panosso Macedo} 
\QMUL
\date{\today}

\begin{abstract}
Motivated by the need of a robust geometrical framework for the calculation of long, and highly accurate waveforms for extreme-mass-ratio inspirals, this work presents an extensive study of the hyperboloidal formalism for the Kerr spacetime and the Teukolsky equation. In a first step, we introduce a generic coordinate system foliating the Kerr spacetime into hypersurfaces of constant time extending between the black-hole horizon and future null infinity, while keeping track of the underlying degrees of freedom. Then, we express the Teukolsky equation in terms of these generic coordinates with focus on applications in both the time and frequency domains. Specifically, we derive a wave-like equation in $2+1$ dimensions, whose unique solution follows directly from the prescription of initial data (no external boundary conditions). Moreover, we extend the hyperboloidal formulation into the frequency domain. A comparison with the standard form of the Teukolsky equations allows us to express the regularisation factors in terms of the hyperboloidal degrees of freedom. In the second part, we discuss several hyperboloidal gauges for the Kerr solution. Of particular importance, this paper introduces the minimal gauge. The resulting expressions for the Kerr metric and underlying equations are simple enough for eventual (semi)-analytical studies. Despite the simplicity, the gauge has a very rich structure as it naturally leads to two possible limits to extremality, namely the standard extremal Kerr spacetime and its near-horizon geometry. When applied to the Teukolsky equation in the frequency domain, we show that the minimal gauge actually provides the spacetime counterpart of the well-known Leaver's formalism. Finally, we recast the hyperboloidal gauges for the Kerr spacetime available in the literature within the framework introduced here.

 \end{abstract}

\maketitle
\section{Introduction}
Black-hole perturbation theory has always been a key topic in the theoretical understanding of general relativity (GR) from both the mathematical and physical perspective. With the dawn of a new era in gravitational-wave (GW) astronomy, observational data of the merger of binary systems~\cite{LIGOScientific:2018mvr} provide the complete picture to the field. The detected wave signals not only enhance our comprehension of astrophysical black holes, but they also raise new challenging questions to contemporary fundamental physics~\cite{Barack:2018yly}. In the near future, testing GR in even stronger conditions shall become routine due to the design/development of the third generation of ground-based detectors, together with the launching of the space-based gravitational wave observatory LISA.

Critical sources of GWs for the LISA Mission are the so-called extreme mass ratio inspirals (EMRI)~\cite{AmaroSeoane2007}, i.e., relatively light objects (e.g., stellar black holes and neutron stars) orbiting a supermassive black hole. They are likely to be found in the centres of galaxies, and the successful detection of their signals brings not only information about the formation and evolution of supermassive black holes~\cite{merritt2013dynamics}, but it also allows us to test Einstein's theory in the strongest gravity regime~\cite{Gair2012nm,Berti2015,Sotiriou2017,Berti2018a,Berti2018b}. Therefore, highly accurate waveform models are crucial to maximising the scientific gain from the GW observations. 

Among the methods to tackle the two-body problem in GR, the gravitational self-force (GSF)~\cite{Detweiler:2005kq,Barack:2009ux,Poisson:2011nh,Wardell:2015kea,Pound:2015tma,Barack:2018yvs} approach is probably the best option to describe EMRI's within the accuracy demanded by LISA. In particular, the optimal parameter estimation within the LISA mission requires  the information coming from the second-order expansion in the binary mass ratio parameter~\cite{Rosenthal:2005ju,Rosenthal:2005it,Rosenthal:2006nh,Rosenthal:2006iy,Detweiler:2011tt,Pound:2012nt,Pound:2012dk,Gralla:2012db,Pound:2014xva}.

Motivated by this enterprise, this paper focuses on a particular geometrical aspect of black-hole perturbation theory, namely the most appropriate choice of coordinates (on a fixed background) to best describe the (infinitely) far away wave zone. For instance, the GSF approach relies on the construction of a {\em retarded} potential, which comes about after fixing boundary conditions to the underlying GSF equations. The boundary conditions must be chosen to describe a physical scenario composed of a black-hole horizon and a radiation zone infinitely far away from the source.  The vast majority of GSF applications use spherical-like coordinates $(t,r,\theta, \varphi)$ with {\em external} boundary conditions imposed in terms of the retarded $u\sim t - r$ or advanced time $v\sim t+r$. As a consequence, Pound points out~\cite{Pound:2015wva} that, at second order, the failure at late times in the GSF programme translates to a failure at large distances as well. Therefore, it has been recently emphasised the need of a robust and systematic framework for adapting the time coordinate to the geometrical structure of the spatial scales near the black hole, and (infinitely) far out at the radiation zone.

\medskip
Formal methods to treat gravitational radiation at large scales take into account the underlying conformal structure of the spacetime~\cite{Penrose:1962ij,Frauendiener2000,Frauendiener02,Kroon:2016ink}. Typically, the physical spacetime $({\cal M}, g_{ab})$ is mapped into a compact, conformal spacetime $(\tilde{\cal M}, \tilde{g}_{ab})$ via $\tilde{g}_{ab} = \Omega^{2} g_{ab}$. The conformal metric is regular, the region in the conformal spacetime where the conformal factor $\Omega$ vanishes capturers the notion of spacetime infinity. In this way, the future endpoints of null geodesics, i.e., the future null infinity $\scri^+$, formally identify the (infinitely) far way radiation zone. Within this framework, one also defines the past null infinity $\scri^-$ as the past endpoints of null geodesics, while space-like infinity $i^0$ follows from $i^0 = \scri^+ \cap \scri^-$. On a (stationary) black-hole spacetime, one identifies further a future horizon ${\cal H}^+$ (the black-hole horizon), a past horizon ${\cal H}^-$ (the white-hole horizon), and the bifurcation sphere ${\cal B} = {\cal H}^+ \cap {\cal H}^-$ (see for instance~\cite{HawEll73}). A practical way to achieve the conformal compactification is via an appropriate choice of coordinates. 

In the standard approaches to black-hole perturbation theory, the preference for spherical-like coordinates $(t,r,\theta, \varphi)$ lies on the simplicity of the resulting equations. However, the limiting regions as $r\rightarrow \infty$ and $r\rightarrow r_{\rm horizon}$ along time surfaces $t=$ constant are precisely the space-like infinity $i^0$ and the bifurcation sphere ${\cal B}$, respectively. Loosely and intuitively speaking, $i^0$ entails features of both future/past null infinity. Similarly, ${\cal B}$ entails features of both future/past horizon. Hence the need for external boundary conditions to fix the physical scenario as composed of a black hole (${\cal H}^+$) and an infinitely far away wave zone ($\scri^+$).

An alternative approach is to introduce a new coordinate system that reaches directly $\scri^+$ as $r\rightarrow \infty$, and eventually the black-hole horizon ${\cal H}^+$ as $r\rightarrow r_{\rm horizon}$. Space-like hypersurfaces of constant time in such system are called (horizon-penetrating) hyperboloidal slices, as they resemble hyperbolas in flat spacetime. This strategy removes the necessity of imposing external boundary conditions since the time coordinate is --- by construction --- naturally adapted to the causal structure of the black hole and the wave zone.

The idea to exploit the freedom in the coordinate choice to reach $\scri^+$ was first put forward in ref.~\cite{Schmi93}, while the more recent ref.~\cite{Zenginoglu:2011jz} argues in favour of the hyperboloidal approach to black-hole perturbation theory. In the last decade, a few choices of hyperboloidal coordinates on fixed backgrounds have been proposed, initially applied to the development of numerical codes and the study of the late time decay of several fields propagating in black-hole spacetimes~\cite{Zenginoglu:2008wc,Zenginoglu:2008uc,Zenginoglu:2009hd,Zenginoglu:2009ey,Bizon:2010mp,Zenginoglu:2010zm,Zenginoglu:2010cq,Racz:2011qu,Jasiulek:2011ce,Harms:2013ib,Yang:2013uba,Spilhaus:2013zqa,Macedo2014,Hilditch:2016xzh,Csukas:2019kcb}. Then, the framework led to some initial studies of EMRI's, mainly within the so-called effective-one-body approach~\cite{Bernuzzi:2010xj,Zenginoglu:2011zz,Bernuzzi:2011aj,Bernuzzi:2012ku,Harms:2014dqa,Nagar:2014kha,Harms:2015ixa,Harms:2016ctx,Lukes-Gerakopoulos:2017vkj}. In the context of the GSF approach, the hyperboloidal formulation was fundamental for the calculation of worldline convolutions~\cite{Zenginoglu:2012xe,Wardell:2014kea}, and it has been recently used in the effective source approach~\cite{Thornburg:2016msc}.

Interestingly, all the works mentioned above treat the underlying equations in the time domain, even though Zenginoglu argued rather early that the hyperboloidal formulation should also lead to efficient codes in the frequency domain~\cite{Zenginoglu:2011jz}. One likely reason is that the coordinates employed in the majority of the studies so far are rather lengthy~\cite{Zenginoglu2018}, thus hindering an early development of (semi-)analytical tools in the frequency domain. The scenario changed with the recent identification of the so-called minimal gauge~\cite{Ansorg2016,Macedo2018} for static black-hole spacetimes. 

With a focus on the Schwarzschild~\cite{Ansorg2016} and Reissner-Nordstr\"om~\cite{Macedo2018} spacetimes, we showed that the hyperboloidal formulation in the minimal gauge provides the geometrical, spacetime counterpart of the well-known Leaver's approach in the frequency domain~\cite{Leaver85,Leaver90}. In particular, we noticed that the minimal gauge provides two limiting processes to extremality, one leading to the usual extremal Reissner-Nordstr\"om black hole, while the second showing a discontinuous transition to the Reissner-Nordstr\"om near-horizon geometry~\cite{Carroll09,Bengtsson:2014fha}. It turns out that the hyperboloidal counterpart of Leaver's approach corresponds to the second formulation. Besides, due to the rather simple analytical structure of the spacetime metric and the wave equations involved, refs.~\cite{Ansorg2016,Macedo2018} expand on Leaver's approach in the frequency domain. The works provide novel tools --- based on the so-called  ``discrete Green's function" technique for recurrence relations --- to express the solutions to wave-like equations in terms of a discrete (quasi-normal-modes) + continuous (tail decay) spectra on non-rotating black-hole spacetimes.

This paper extends the geometrical results from refs.~\cite{Ansorg2016,Macedo2018} into the Kerr spacetime. Here, we introduce a generic formulation for the hyperboloidal approach of the Kerr spacetime and its perturbation equations. The work provides the theoretical tools for a robust hyperboloidal framework, which will serve as the basis for further studies of EMRI's focusing on the production of highly accurate waveform signals templates for the data analysis pipeline in the LISA Mission. Specifically, sec. \ref{sec:Kerr} introduces the generic hyperboloidal coordinate system for the Kerr spacetime and scrutinises the degrees of freedom within the coordinate choice. Then, sec.~\ref{sec:TeukEq} develops the hyperboloidal formulation of the Teukolsky equation in both time and frequency domain. Finally, secs.~\ref{sec:MG} and ~\ref{sec:HypGauges} discuss several choices of hyperboloidal gauges for the Kerr spacetime. The former focuses on the simplest hyperboloidal foliation for the Kerr spacetime --- the minimal gauge. The latter reviews and re-casts all hyperboloidal gauges available in the literature in terms of the formalism introduced here.

This work uses natural units where $G=c=1$.

\section{Kerr spacetime}\label{sec:Kerr}
In this section, we present the generic formalism for the construction of hyperboloidal slices on a fixed background with the focus on the Kerr spacetime.
 
\subsection{Boyer-Lindqst coordinates}
We begin by reviewing the Kerr spacetime in Boyer-Lindqst coordinate $(t,r,\theta, \varphi)$
\bea
\label{eq:Kerr_BL}
\d s^2 &=& -f\d t^2 - \dfrac{4Mar}{\Sigma}\sin^2\theta \d t \d\varphi + \dfrac{\Sigma}{\Delta} \d r^2 + \Sigma \d\theta^2 \nn \\
&& + \sin^2\theta\left( \Sigma_0 + \dfrac{2Ma^2r}{\Sigma}\sin^2\theta \right)\d\phi^2,
\eea
with
\bea
&&\Delta(r) =  r^2 -2Mr + a^2 = \bigg(r-r_+\bigg)\bigg(r-r_-\bigg), \label{eq:Delta}  \\
&&\Sigma(r,\theta) = r^2 + a^2\cos^2\theta, \quad
\Sigma_0(r) = \Sigma(r,0) = r^2 + a^2, \label{eq:Sigmas}\\ 
&&f(r,\theta) = 1 - \frac{2Mr}{\Sigma(r,\theta)}. \label{eq:f}
\eea
As usual, the parameters $M$ and $a$ relate, respectively, to the black hole's mass and angular momentum. The condition $\Delta(r) = 0$ defines the event ($r_+$) and Cauchy ($r_-$) horizons
\bea
r_\pm = M\left( 1 \pm \sqrt{1 - \frac{a^2}{M^2}}\right).
\eea
It will be convenient to parametrise the spacetime via $\kappa\in[-1,1]$ defined by
\bea
\label{eq:KerrParm}
\kappa^2:= \frac{r_-}{r_+}  \Rightarrow r_+ = \frac{2M}{1+\kappa^2}, \quad r_- = \frac{2M}{1+\kappa^2} \kappa^2, \quad a = r_+ \kappa.
\eea
In terms of the usual dimensionless spin parameter $j = a/M$, it reads
\beq
\kappa = \dfrac{j}{1+\sqrt{1-j^2}} \label{eq:kappa_a_over_M}.
\eeq
Thus, $\kappa=0$ ($a=0$) reduces the metric to the Schwarzschild spacetime while $|\kappa| = 1$ ($|a| = M$) leads to the extremal Kerr solution. 

Along the hypersurfaces $t=$ constant, the limit $r\rightarrow \infty$ leads to spatial infinity $i^0$, whereas $r=r_+$ corresponds to the bifurcation sphere.

\subsection{Ingoing Kerr coordinates}\label{sec:KS}
Since we wish to construct horizon penetrating coordinates, we first introduce the ingoing Kerr coordinates $(v,r,\theta,\phi)$ via
\beq
\label{eq:TrasnfBL_KS}
t = v - r^*(r), \quad \varphi = \phi - k(r), 
\eeq
with the tortoise coordinate $r^*(r)$ and the phase $k(r)$ defined by
\beq
\dfrac{d r^*}{dr} = \dfrac{\Sigma_0}{\Delta},  \quad \dfrac{d k}{dr} = \dfrac{a}{\Delta}.
\eeq
Note that the tortoise coordinate $r^*$ and the phase $k$ are defined up to an overall constant. When needed, we use in this work 
\bea 
r^* &=& r + \dfrac{2 M}{1-\kappa^2}\ln\left(\dfrac{r}{r_+}-1\right) - \kappa^2 \dfrac{2M}{1-\kappa^2}\ln\left(\dfrac{r}{r_+}-\kappa^2\right) \\  
k &=& \dfrac{\kappa}{1-\kappa^2} \ln\left( \dfrac{r/r_+ - 1}{ r/r_+ - \kappa^2 } \right).
\eea
The line element~\eqref{eq:Kerr_BL} then transforms into its original Kerr's form
\bea
\label{eq:Kerr_KSCoord}
\d s^2 &=& - f \bigg( \d v - a\sin^2\theta \d \phi \bigg)^2 + \Sigma \d\omega^2 \nn \\ &&+ 2 \bigg( \d v - a \sin^2\theta \d\phi \bigg)\bigg( \d r - a \sin^2\theta \d\phi \bigg) ,
\eea
with $\d\omega^2 = \d\theta^2 + \sin^2\theta \d\phi$ the line element of the unit sphere. 
By construction, the surface $r=r_+$ along $v=$ constant corresponds to the future black-hole horizon. However, the limit $r\rightarrow \infty$ leads to past null infinity $\scri^-$. 

In the next section, we introduce coordinates for which $r\rightarrow \infty$ corresponds, actually, to future null infinity $\scri^+$. Following~\cite{Macedo2018}, to identify $\scri^+$ it is convenient to keep track of the null vectors $k^a$ and $l^a$ associated to the ingoing and outgoing light rays, respectively. In the present coordinate system, they read
\beq
\label{eq:NullVec}
k^a = -\zeta \delta^a_r, \quad l^a = \zeta^{-1} \Bigg( \frac{r^2+a^2}{\Sigma}\delta^a_v + \frac{\Delta}{2\Sigma}\delta^a_r + \frac{a}{\Sigma}\delta^a_\phi \Bigg).
\eeq
The (free) boost parameter $\zeta$ is fixed in the next section. 

\subsection{Conformal compactfication and the hyperboloidal slicing}
We finally introduce compact hyperboloidal coordinates $(\tau, \sigma,\theta,\phi)$ via the height function technique~\cite{Zenginoglu:2007jw}. Extending on~\cite{Macedo2018}, we consider a $\theta$-dependence via
\bea
\label{eq:Coords_Hyp}
v = \lambda \bigg( \tau - h(\sigma,\theta) \bigg), \quad r = \lambda \frac{\rho(\sigma)}{\sigma},
\eea
with $\lambda$ a length scale of the spacetime. \ref{App:LengthScale} discusses different choices for $\lambda$. The height function $h(\sigma,\theta)$ and the radial function $\rho(\sigma)$ encode the gauge degrees of freedom. The radial compactification allow us to naturally associate a conformal factor in terms of the new coordinate $\sigma$ via 
\beq
\label{eq:ConfFact}
\Omega = \sigma/\lambda,
\eeq 
which leads to a conformal spacetime with line element\footnote{\ref{App:Metric} brings explicitly the components of the metric, its inverse and determinant.
}
\bea
\d\tilde s^2 &=& \Omega^2 \d s^2  \nn \\
&=& -\sigma^2 F \Bigg(\d\tau - h_{,\sigma} \d\sigma - h_{,\theta}\d\theta - \alpha\sin^2\theta \d\phi \Bigg)^2 + \tilde\Sigma\d\omega^2 \label{eq:ConformalKerr} \\
&& -2\Bigg(\d\tau - h_{,\sigma} \d\sigma - h_{,\theta}\d\theta - \alpha\sin^2\theta \d\phi \Bigg) \Bigg(\beta\d\sigma + \alpha\sigma^2\sin^2\theta\d\phi \Bigg). \nn
\eea
In the above expression, the gauge freedom in the radial direction is captured by the shift
\beq
\label{eq:beta}
\beta(\sigma) = \rho(\sigma) - \sigma \rho'(\sigma).
\eeq
Moreover, eqs.~\eqref{eq:Delta}-\eqref{eq:f} transform under the coordinate change and the conformal compactification to  
\bea
\tilde\Delta(\sigma) = \Omega^{2}\Delta(r(\sigma)) &=& \rho(\sigma)^2 - 2\mu\rho(\sigma)\sigma + \alpha^2\sigma^2  \nn \\
&=& \Bigg( \rho(\sigma) - \dfrac{r_+}{\lambda} \sigma \Bigg)\Bigg( \rho(\sigma) - \dfrac{r_-}{\lambda} \sigma \Bigg). \label{eq:ConfResDelta} \\
\tilde\Sigma(\sigma,\theta) = \Omega^2\, \Sigma(r(\sigma),\theta) &=& \rho(\sigma)^2 + \alpha^2\,\sigma^2 \cos^2\theta, \\
\tilde\Sigma_0(\sigma) = \Omega^{2}\Sigma_0(r(\sigma)) &=& \rho(\sigma)^2 + \alpha^2\sigma^2, \\
F(\sigma,\theta) = f(r(\sigma), \theta) &=& 1 - \frac{2\,\mu\,\rho(\sigma)\,\sigma}{\tilde\Sigma(\sigma,\theta)},
\eea
with the dimensionless mass and spin parameters respectively (cf.~\ref{App:LengthScale})
\beq 
\label{eq:MassSpinPar}
\mu=M/\lambda, \quad \alpha=a/\lambda.
\eeq 

\medskip
The conformal null vectors re-scale as $\tilde{k}^a = \Omega^{-1}\, k^a$ and $\tilde{l}^a = \Omega^{-1}\, l^a$. To ensure that the hypersurfaces $\tau=$ constant foliate future null infinity, we require $\tau$ to be a good parameter of the ingoing conformal null vector via 
$ \tilde{k}^a \partial_a \tau = 1$. This requirement fixes the boost parameter $\zeta$ in~\eqref{eq:NullVec}, and it leads to
\bea
\tilde k^a &=& \delta^a_\tau + \frac{1}{h_{,\sigma}}\delta^a_\sigma, \\
 \tilde l^a &=& \frac{h_{,\sigma}}{2\beta^2\tilde\Sigma}\left( 2\beta\tilde\Sigma_0- \tilde\Delta\sigma^2h_{,\sigma}   \right)\delta^a_\tau - \frac{\tilde\Delta\sigma^2h_{,\sigma}}{2\beta^2\tilde\Sigma}\delta^a_\sigma +\frac{ \alpha \sigma^2 h_{,\sigma}}{\beta\tilde\Sigma}\delta^a_\phi. \label{eq:ConfNullTetrad}
\eea
Finally, we impose that $\sigma=0$ is a null surface corresponding to future null infinity via~\cite{Macedo2018}
\beq
\lim_{\sigma\rightarrow 0}\tilde k^a = \delta^a_\tau \Rightarrow \lim_{\sigma\rightarrow 0} \frac{1}{h_{,\sigma}} = 0.
\eeq
However, the above condition must not jeopardise the regularity of the outgoing conformal null vector $\tilde l^a$ as $\sigma\rightarrow 0.$

We recall that for non-rotating black holes, $\rho(\sigma)$ is the areal radius in the conformal representation of the spacetime. As such, it was natural to considered it to be a regular function on its domain attaining positive, non-vanishing values\cite{Macedo2018}. We assume the same properties here. If one considers
\beq
\rho(\sigma) = \rho_0 + \sigma\rho_1 + {\cal O}(\sigma^2) \Rightarrow \beta(\sigma) = \rho_0 + {\cal O}(\sigma^2), \label{eq:AsympExp_rho_beta}
\eeq
and expands all the relevant quantities around $\sigma=0$, one obtains the same result as in~\cite{Macedo2018}, i.e., the spin parameter of the Kerr solution does not affect the leading terms in the height function. Specifically, the components of $\tilde l^a$ remain finite for
\beq
h_{,\sigma} = \frac{2\rho_0}{\sigma^2}\left[ 1 + \frac{2\mu}{\rho_0}\sigma\right] + {\cal O}(1) \Longrightarrow  h(\sigma,\theta) = h_0(\sigma) +A(\sigma,\theta), \label{eq:Height}
\eeq
with
\beq
 h_0(\sigma) = -2\rho_0 \left[ \frac{1}{\sigma}-\frac{2\mu}{\rho_0}\ln\sigma\right] \label{eq:HeightMin}.
\eeq
The time function $A(\sigma,\theta)$ together with the radial shift $\beta(\sigma)$ account for all gauge degrees of freedom. The only restriction on their choice is that the surfaces $\tau=$ const. are spacelike outside the black-hole region. Thus, $\tilde\nabla_a\tau \tilde\nabla^a\tau < 0$ imposes 
\beq
y_-< y < y_+,
\eeq
with
\beq
y:= -\lambda \frac{d}{dr^*}h = \frac{\sigma^2 \tilde\Delta h_{,\sigma}}{\tilde\Sigma_0\beta}, \quad
y_{\pm} = 1 \pm \sqrt{1 - \frac{\tilde\Delta \sigma^2}{\tilde\Sigma_0^2}\left(h_{,\theta}^2 + \alpha^2\sin^2\theta\right)}.
\eeq

\section{Teukolsky Equation}\label{sec:TeukEq}
The (sourceless) Teukolsky equation (TE) --- originally derived in the Boyer-Lindqst coordinates $x^a=(t,r,\theta,\varphi)$ --- reads for the master function $\Psi(x^a)$
\bea
&0=\left[  \dfrac{(\Sigma_0)^2}{\Delta} - a^2 \sin^2\theta \right] \partial^2_{tt} \Psi + \dfrac{4Mar}{\Delta} \partial^2_{t\varphi} \Psi + \left[ \dfrac{a^2}{\Delta} - \dfrac{1}{\sin^2\theta} \right] \partial^2_{\varphi\varphi}\Psi  \nn \\
&    - \Delta^{-p}\partial_r \bigg( \Delta^{p+1} \partial_r \Psi \bigg)  - 2p\left[ \dfrac{M(r^2-a^2)}{\Delta} - (r+ia\cos\theta) \right] \partial_t\Psi \label{eq:Teukolsky_BL}   \\
&- 2p \left[ \frac{a(r-M)}{\Delta}  + i\dfrac{\cos\theta}{\sin^2\theta}\right]\partial_\varphi \Psi - \dfrac{1}{\sin\theta}\partial_\theta \bigg( \sin\theta \partial_\theta \Psi\bigg) + p(p\cot^2\theta -1)\Psi. \nn
\eea
Here, $p$ is the field's spin-weight parameter and it describes scalar ($p=0$), electromagnetic ($p=\pm 1$) and gravitational ($p=\pm 2$) perturbation. Eq.~\eqref{eq:Teukolsky_BL} must be solved with ingoing and outgoing boundary conditions at the horizon $r_+$ and spatial infinity $r \rightarrow \infty$, respectively.

One can separate the Teukolsky  eq.~\eqref{eq:Teukolsky_BL} in the frequency domain~\cite{Teukolsky72,Teukolsky:1973ha}. The ansatz
\beq
\label{eq:ansatzFreq_BL}
\Psi(t,r,\theta,\varphi) = e^{-i\omega\, t} {\cal R}(r) S(\theta) e^{im\varphi}
\eeq
leads to two ordinary differential equations for the functions ${\cal R}(r)$ and $S(\theta)$
\bea
\boldsymbol{ {\cal T}} && S(\theta) = 0, \quad \boldsymbol{ {\cal T}} = \dfrac{1}{\sin\theta} \dfrac{\d}{\d \theta}\left( \sin\theta \dfrac{\d}{\d \theta} \right) + a^2\omega^2\cos^2\theta - 2p a\omega\cos\theta \nn \\
&& -\left(\dfrac{m+p\cos\theta}{\sin\theta}\right)^2+p+A_{\ell m}, \label{eq:SpinWeightSphHarm}\\
\boldsymbol{{\cal D}}&&  \,{\cal R}(r) = 0, \quad \boldsymbol{{\cal D}} = \Delta^{-p}\dfrac{\d}{\d r} \left( \Delta^{p+1} \dfrac{\d }{\d r}  \right) + \left( 2ip\omega r -a^2\omega^2 -A_{\ell m} \right) \label{eq:QNM_BL}  \\
&& +   \dfrac{ (\omega\Sigma_0 )^2 - 4Mam\omega r + a^2m^2+ 2ip\left[ am(r- M)-M\omega(r^2-a^2) \right] }{\Delta}. \nn
\eea
Eq.~\eqref{eq:SpinWeightSphHarm} forms a Sturm-Liouville eigenvalue problem, and the solutions ${}_p S_{\ell m}(\theta)$ are the so-called spin-weighted spheroidal harmonics~\cite{Teukolsky72,Teukolsky:1973ha}. Then, eq.~\eqref{eq:QNM_BL} leads to the quasinormal modes $\omega_n$ when solved with the ingoing (outgoing) boundary conditions at the horizon (spatial infinity), together with appropriate regularity conditions\footnote{Conditions leading to quasinormal modes are often expressed {\em solely} by the asymptotic behaviours of ${\cal R}(r)$ as $r^*\rightarrow \pm \infty$. Though necessary, such conditions are not sufficient --- see e.g.~section~3.1.2 in ref.~\cite{Nollert99}. Ref.~\cite{Ansorg2016} employs numerical tools to address important issues regarding the regularity of the underlying quasinormal mode eigenfunctions, whereas refs.~\cite{Gajic:2019qdd,Gajic:2019oem} brings a rigorous mathematical discussion on the matter.}.
In the next subsections, we discuss the TE in the time and frequency domain within the hyperboloidal formalism introduced in the last section.

\subsection{Time domain}
We aim initially at re-writing eq.~\eqref{eq:Teukolsky_BL} in the hyperboloidal coordinates $\chi^a = (\tau,\sigma,\theta,\phi)$ in terms of a master function $U(\chi^a)$ which is regular at future null infinity $\sigma=0$ and the horizon $\sigma=\sigma_+$. Then, thanks to the axial-symmetry, we introduce a Fourier decomposition in the coordinate $\phi$. With a regularisation at the poles of the spherical coordinates $\sin\theta=0$, the final goal is to obtain an equation in $2+1$ dimensions for a given Fourier mode $V_m(\tau,\sigma,\theta)$ which is regular at the radial and angular boundaries. 

\subsubsection{Regularity at radial boundaries\\}
Initially, one applies the transformation $x^a=x^a(\chi^b)$ from the Boyer-Lindqst to the hyperboloidal coordinates directly to eq.~\eqref{eq:Teukolsky_BL}. After the coordinate change, the regularisation of essential singularities in the radial direction at future null infinity $\sigma=0$ and the black-hole horizon $\sigma=\sigma_{\rm +}$ follows from
\beq
\label{eq:RegField}
U(\chi^a) = \Omega^{-1} \,\Delta^p \, \Psi\bigg( x^b(\chi^a) \bigg).
\eeq
In particular, one can follow the intermediary step for section \ref{sec:KS} and first transform the TE from the Boyer-Lindqst into the ingoing Kerr coordinates --- see eqs.~\eqref{eq:TrasnfBL_KS}. The factor $\Delta^p$ in eq.~\eqref{eq:RegField} enters at this stage to regularise the field at the horizon. Then, one transforms from ingoing Kerr to hyperboloidal coordinates according to eq.~\eqref{eq:Coords_Hyp}. The factor $\Omega^{-1}$ in eq.~\eqref{eq:RegField} guarantees the regularity at $\scri^+$. Indeed, the final form of the TE in the hyperboloidal formulation reads
\bea
&&\Bigg( \dfrac{\tilde\Sigma_0}{\beta}h_{,\sigma}\left[ 2 - \frac{\sigma^2 \tilde\Delta h_{,\sigma}}{\tilde\Sigma_0\beta} \right] - \left[h_{,\theta}^2 + \alpha^2\sin^2\theta\right] \Bigg) U_{,\tau \tau} +{\rm c}_\tau U_{,\tau} +2\alpha\frac{\sigma}{\beta}  U_{,\phi}  \nn \\
&&2\dfrac{\tilde\Sigma_0}{\beta}\left[ 1-\frac{\sigma^2 \tilde\Delta h_{,\sigma}}{\tilde\Sigma_0\beta} \right]U_{,\tau\sigma} - 2\alpha\left[ 1-\dfrac{\sigma^2 h_{,\sigma}}{\beta} \right]U_{,\tau\phi} + 2\alpha \dfrac{\sigma^2}{\beta} U_{,\sigma\phi} -2h_{,\theta}U_{,\tau\theta} \label{eq:Teukolsky_U} \\
&&- \dfrac{\tilde\Delta^p}{\beta \sigma^{2p}} \left[ \dfrac{\sigma^{2(1+p)} \tilde\Delta^{1-p}  }{\beta}  U_{,\sigma}   \right]_{,\sigma} - \left( 
\bar{\eth} \eth
-2p 
+ \frac{\tilde\Delta^p}{\beta}\sigma^{1-2p} \left[ \frac{\sigma^{2p}\tilde\Delta^{1-p}}{\beta}\right]_{,\sigma} 
 \right) U =0 . \nn \eea
The derivatives in the angular directions were incorporated in terms of the eth-operator~\cite{Goldberg67} $\eth$ acting on the field $U(\chi^a)$ with spin weight $p$
\bea
\bar{\eth}\eth U &=&  \dfrac{\partial_\theta \bigg( \sin\theta \partial_\theta U\bigg)}{\sin\theta} + \dfrac{\partial^2_{\phi\phi}U}{\sin^2\theta} + 2i\, p \dfrac{\cos\theta}{\sin^2\theta}\partial_\phi U + p(1-p\cot^2\theta )U.
\eea
Moreover, the coefficient ${\rm c}_\tau$ is given by
\bea
{\rm c}_\tau &=& -\dfrac{\tilde\Delta^p}{\beta \sigma^{2p}} \partial_\sigma \left[ \dfrac{\sigma^{2(1+p)} \tilde\Delta^{1-p}  }{\beta}h_{,\sigma} \right]  + \dfrac{2}{\beta\sigma}\tilde\Sigma_0 - \dfrac{2\rho}{\sigma}(1-2p) \nn \\
&&+ 2i\, p\, \alpha\cos\theta -\dfrac{ \partial_\theta \bigg( \sin\theta h_{,\theta}\bigg)}{\sin\theta}. \label{eq:alpha_tau}
\eea 
Despite its apparent singular behaviour at $\sigma=0$, one verifies that $c_\tau$ actually behaves as
\beq
{\rm c}_\tau \sim \frac{2\rho_0}{\beta^2}\left[ 2\rho_0 - \beta(1-2p)\right]\frac{\beta - \rho_0}{\sigma} \sim {\cal O}(\sigma).
\eeq
The result follows from $\beta(\sigma) -\rho_0 = {\cal O}(\sigma^2) $, together with the height function's leading order behaviour  --- see eqs.~\eqref{eq:AsympExp_rho_beta} and~\eqref{eq:Height}.

Finally, we observe that the particular form of the last terms in eq.~\eqref{eq:Teukolsky_U} leads to 
\beq -2p + \frac{\tilde\Delta^p}{\beta}\sigma^{1-2p} \left[ \frac{\sigma^{2p}\tilde\Delta^{1-p}}{\beta}\right]_{,\sigma}  \sim {\cal O}(\sigma).\eeq Alternatively, the factor $-2p$ could be incorporated into the angular operator via  $\eth\bar{\eth} = \bar{\eth}\eth - 2p$. The chosen option allows for more straightforward comparison with the standard equations in the frequency domain --- see sec.~\ref{sec:HypFreqDom}.

The re-scaling~\eqref{eq:RegField} does not remove the degeneracies of the TE at the two boundaries altogether. In fact, the term proportional to $\partial_{\sigma\sigma} U$ goes as $\sigma^2 \tilde{\Delta}$, which vanishes at  $\sigma=0$ and $\sigma=\sigma_+$. Such degeneracies provide boundary conditions guaranteeing that the characteristics of the wave equation always point outward the numerical domain. Hence, when looking for {\em regular} solutions, no further boundary conditions at the horizon nor future null infinity are allowed to be imposed.


\subsubsection{Conformal re-scaling and peeling properties\\}
The peeling theorem for the Newman-Penrose scalars states that\footnote{Quantities denoted with a bar represent a regular function of order ${\cal O}(1)$ as $\Omega\rightarrow 0$. They may not necessarily coincide with the conformal Newman-Penrose quantities derived directly from the conformal metric \eqref{eq:ConformalKerr} and the tetrads \eqref{eq:ConfNullTetrad}.}
\beq
\Psi_k = \Omega^{5-k}\bar\Psi_k, \quad  (k=0\cdots 4); \quad \phi_n = \Omega^{3-n}\bar \phi_n, \quad (n=0\cdots 2), \nn
\eeq
while the scalar field transform as $\Phi = \Omega \bar \Phi$. Apart from that, the Teukolsky  master function $\Psi$ includes a pre-factor $\varrho^{2p}$ whenever $p<0$, with $\varrho =-( r +i\, a\cos\theta)^{-1}$ the Newman-Penrose spin coefficient. Its direct re-scaling reads
\beq
\varrho= \Omega \bar\varrho, \quad \bar\varrho = - \bigg( \rho(\sigma) + i\, \alpha \cos\theta\bigg)^{-1}. \nn
\eeq
By systematically collecting all the conformal factors for the master functions with different spins, one obtains the asymptotic behaviour $\Psi \sim \Omega^{1+2p}$ --- see table~\ref{tab:FieldPert}. Such asymptotic behaviour follows straightforwardly from eq.~\eqref{eq:RegField}. Indeed, with the conformal re-scaling of the function $\Delta$ in eq.~\eqref{eq:ConfResDelta}, eq.~\eqref{eq:RegField} yields
\beq
\label{eq:ConformalMasterFunction}
U \sim \Omega^{-(1+2p)} \Psi \Longrightarrow U \stackrel{\Omega\rightarrow 0}= {\cal O}(1).
\eeq
\begin{table}[h]
\caption{Master function and its asymptotic behaviour}
\begin{center}
\begin{tabular}{|c|c|c|}
\hline
$\quad p \quad$ & $\quad \Psi \quad$ & $\Omega\rightarrow 0$ \\
\hline
\hline
-2 & $\quad \varrho^{-4}\Psi_4 $& ${\cal O}( \Omega^{-3} )$ \\
\hline
-1 & $\quad \varrho^{-2}\phi_2$& ${\cal O}( \Omega^{-1} )$ \\
\hline
0 & $\quad \Phi$& ${\cal O}( \Omega )$ \\
\hline
1 & $\quad \phi_0$& ${\cal O}( \Omega^3 )$ \\
\hline
2 & $\quad \Psi_0$& ${\cal O}( \Omega^5 )$ \\
\hline
\end{tabular}
\end{center}
\label{tab:FieldPert}
\end{table}

\subsubsection{Evolution equation in $2+1$ dimensions \\}
Finally, we exploit the axial-symmetry of the system to decompose the solution into its Fourier modes
\beq
U(\tau,\sigma,\theta,\phi) = \sum_{m=-\infty}^{\infty}U_{m}(\tau, \sigma, \theta) e^{im\phi}.
\eeq
To regularise the essential regularity at $\sin(\theta) = 0$, we introduce one last transformation
\beq
\label{eq:ReScaleUm_to_Vm}
U_m(\tau,\sigma,\theta) = \cos^{\delta_1}(\theta/2) \sin^{\delta_2}(\theta/2) V_m(\tau,\sigma,\theta),
\eeq
with the exponents $\delta_1 = \left| m-p \right|$ and $\delta_2 = \left| m+p \right|$. 

With the substitution $x=\cos\theta$ one achieves the final (regular) form for the Teukolsky  equation\footnote{In eq.~\eqref{eq:Teukosly_Vm} and further equations in this section, we abuse the notation for the functions regarding the substitution $x=\cos\theta$ and consider the notation $f(\theta)$ as resulting from $f(x(\theta))$.}
\bea
&&0=\Bigg( \dfrac{\tilde\Sigma_0}{\beta}h_{,\sigma}\left[ 2 - \frac{\sigma^2 \tilde\Delta h_{,\sigma}}{\tilde\Sigma_0\beta} \right] - (1-x^2)\left[h_{,x}^2 + \alpha^2\right] \Bigg) V_m{}_{,\tau\tau} -  \left[ (1-x^2) V_m{}_{,x}\right]_{,x}    \nn \\
&& + 2i\, \alpha m \dfrac{\sigma^2}{\beta} V_m{}_{,\sigma} +2\dfrac{\tilde\Sigma_0}{\beta}\left[ 1-\frac{\sigma^2 \tilde\Delta h_{,\sigma}}{\tilde\Sigma_0\beta} \right]V_m{}_{,\tau\sigma} -2(1-x^2)h_{,x}\,V_m{}_{,\tau x}  \nn\\
&& + \left[ {\rm c}_\tau  - 2i\, \alpha m\left( 1-\dfrac{\sigma^2 h_{,\sigma}}{\beta} \right) + \left[ (1-x)\delta_1 - (1+x) \delta_2   \right]  h_{,x}  \right] V_m{}_{,\tau}   \label{eq:Teukosly_Vm} \\
&&- \dfrac{\tilde\Delta^p}{\beta \sigma^{2p}} \left[ \dfrac{\sigma^{2(1+p)} \tilde\Delta^{1-p}  }{\beta} V_m{}_{,\sigma}   \right]_{,\sigma}  
  + \left[ (1+x) \delta_2 - (1-x)\delta_1 \right]  V_m{}_{,x}  + \left[  \frac{2i \alpha m\sigma}{\beta} \right.  \nn \\
&& \left.  - \frac{\sigma^{1-2p}\tilde\Delta^p}{\beta} \left( \frac{\sigma^{2p}\tilde\Delta^{1-p}}{\beta}\right)_{,\sigma} + 2p  - \left( p - \dfrac{\delta_1+\delta_2}{2} \right)\left(1+p+\dfrac{\delta_1+\delta_2}{2}\right) \right] V_m . \nn
\eea
Similar to the behaviour at the boundaries in the radial direction, the equation still degenerates at $x=\pm1$ due to the vanishing of the coefficient in front of $\partial^2_{xx} V_m$. No extra boundary conditions are needed thereon as the corresponding regularity conditions at the north and south poles of the spherical-type coordinate system must be imposed. 

Eq.~\eqref{eq:Teukosly_Vm} provides a hyperbolic equation for the regular field $V_m(\tau,\sigma,x)$ defined in the domain $(\tau,\sigma,x)\in[\tau_0,\tau_1]\times[0,1]\times[-1,1]$. A unique time evolution follows from initial data
\beq
V_{m}^0(\sigma, x) = V_m(\tau_0, \sigma, x), \quad W_{m}^0(\sigma, x) = \partial_\tau V_m(\tau_0, \sigma, x).
\eeq

\subsection{Frequency Domain}\label{sec:TeukEq_FreqDomain}
We now approach the hyperboloidal formulation of TE in the frequency domain. The most straightforward strategy is to directly Fourier-transform eq.~\eqref{eq:Teukosly_Vm} into the frequency domain. This procedure allows us to identify the key elements within the frequency domain for the direct hyperboloidal transformation of the eqs.~\eqref{eq:SpinWeightSphHarm} and \eqref{eq:QNM_BL}.

\subsubsection{Fourier-transform of hyperboloidal time\\}
We begin by considering the Ansatz
\bea
V_m(\tau,\sigma,\theta) = e^{s\tau} {\rm v}(\sigma)\, {\mathbb S}(\theta) \label{eq:ansatzFreq_Hyp_Vm}.
\eea
To enable a separation of variables, the height function must decompose as
\beq
\label{eq:HeightDecomp}
h(\sigma,\theta) = H_0(\sigma) + H_1(\theta).
\eeq
At this point, it is essential to recall eqs.~\eqref{eq:Height} and \eqref{eq:HeightMin}, which identifies the irregular leading term $h_0(\sigma)$ and the regular function $A(\sigma,\theta)$. While the former yields the desired hyperboloidal behaviour for the foliation, the latter captures the gauge degrees of freedom. Thus, in the decomposition~\eqref{eq:HeightDecomp}, the radial dependence $H_0(\sigma)$ must not necessarily coincide with the leading term $h_0(\sigma)$. It is actually the regular part $A(\sigma, \theta)$ that must be separable leading to $H_0(\sigma) = h_0(\sigma) + A_0(\sigma)$ and $H_1(\theta) = A_1(\theta)$.

Eq.~\eqref{eq:Teukosly_Vm} leads to the following ordinary differential equations for the functions ${\mathbb S}(x)$ and ${\rm v}(\sigma)$
\bea
&&\boldsymbol{\Theta} {\mathbb S}(x) = 0, \quad  \quad \boldsymbol{\Theta}=   (1-x^2) \dfrac{\d^2}{\d x^2} \nn \\
&& + \bigg(2s(1-x^2)H_1' -2x  -  (1+x) \delta_2 + (1-x)\delta_1  \bigg)\dfrac{\d}{\d x}  \label{eq:SpinWeightSphHarm_Hyp} \\
&& +\Bigg(A_{\ell m} + \left( p - \dfrac{\delta_1+\delta_2}{2} \right)\left(1+p+\dfrac{\delta_1+\delta_2}{2}\right) + \left[ (1-x^2)H_1'{}^2 - \alpha^2 x^2\right]  s^2 \nn \\
&& - s\bigg[  2i\, p\, \alpha x - \bigg( (1-x^2) H_1'\bigg)_{,x} - \left[ (1-x)\delta_1 - (1+x) \delta_2   \right]  H_1' \bigg]  \Bigg),  \nn 
\eea
\bea
&& \boldsymbol{A} {\rm v}(\sigma) = 0, \quad \quad  \boldsymbol{A} = \dfrac{\tilde\Delta^p}{\beta \sigma^{2p}}\dfrac{\d}{\d \sigma} \left[ \dfrac{\sigma^{2(1+p)} \tilde\Delta^{1-p}  }{\beta} \dfrac{\d}{\d \sigma}    \right] \nn \\
 &&   -2\Bigg( s\dfrac{\tilde\Sigma_0}{\beta}\left[ 1-\frac{\sigma^2 \tilde\Delta H_0'}{\tilde\Sigma_0\beta} \right] + i \alpha m \dfrac{\sigma^2}{\beta} \Bigg)  \dfrac{\d}{\d \sigma}   - 
  \Bigg( \dfrac{\tilde\Sigma_0H_0'}{\beta}\left[ 2 - \frac{\sigma^2 \tilde\Delta H_0'}{\tilde\Sigma_0\beta} \right] - \alpha^2\Bigg)s^2 \nn\\
  && -s \left[-\dfrac{\tilde\Delta^p}{\beta \sigma^{2p}} \dfrac{\d}{\d\sigma}\left( \dfrac{\sigma^{2(1+p)} \tilde\Delta^{1-p} H_0'  }{\beta}  \right)    - \dfrac{2\rho(1-2p)}{\sigma}  - 2i\alpha m\left( 1-\dfrac{\sigma^2}{\beta}H_0' \right) \right.       \nn \\
 && \left. + \dfrac{2\tilde\Sigma_0}{\beta\sigma} \right] +  \frac{2i\, \alpha m\sigma}{\beta}  + \frac{\tilde\Delta^p\sigma^{1-2p}}{\beta} \left( \frac{\sigma^{2p}\tilde\Delta^{1-p}}{\beta}\right)_{,\sigma} - 2p - A_{\ell m}. \label{eq:QNM_Hyp} 
\eea
Eqs.~\eqref{eq:SpinWeightSphHarm_Hyp} and \eqref{eq:QNM_Hyp} inherit the degeneracies of eq.~\eqref{eq:Teukosly_Vm}. In particular, eq.~\eqref{eq:SpinWeightSphHarm_Hyp} presents a regular singular point at $x=\pm1$,  which allows one to seek solutions in the form
\beq
\label{eq:Taylor_S}
{\mathbb S}(x) = \sum_{n=0}^{\infty} {\rm a}_n (1+x)^n.
\eeq
Whenever $H_1(x)$ is polynomial in $x$, the Ansatz \eqref{eq:Taylor_S} leads to a recurrence relation for the coefficients ${\rm a}_n$. In this case, the methods developed in refs.~\cite{Leaver85,Macedo2018} can be used to find the angular eigenvalues $A_{\ell m}$ and to construct the (re-scaled) spin weighted spherical harmonics ${}_p{\mathbb S}_{\ell m}(x)$. For instance, Leaver's well-known $3-$term recurrence relation~\cite{Leaver85} follows from --- see section~\ref{sec:NewmanJanis_Leaver} for further comments on this choice
\beq
\label{eq:LeaverAngHeight}
H_1(x) = i \alpha \, x.
\eeq
We observe that the above choice is not essential to the study of eq.~\eqref{eq:SpinWeightSphHarm_Hyp} with the Ansatz \eqref{eq:Taylor_S}. The option $H_1(x) = 0$ can also be employed, but it leads to a $4$-term recurrence relation for the coefficients ${\rm a}_n$.

Eq. \eqref{eq:QNM_Hyp}, has a regular singular point at $\sigma=\sigma_+$ and one can express the solution as
\beq
\label{eq:Taylor_v}
{\rm v}(\sigma) = \sum_{n=0}^{\infty} {\rm b}_n \left(1-\dfrac{\sigma}{\sigma_+}\right)^n.
\eeq
Similar to the angular equation, whenever $H_0(\sigma)$ yields an eq.~\eqref{eq:QNM_Hyp} whose coefficients are polynomial in $\sigma$, a recurrence relation for ${\rm b}_n$ follows. The quasi-normal frequencies $s_{\ell m}$ are obtained for ${\rm v}(\sigma)$ sufficiently regular in the entire domain $\sigma\in[0,\sigma_+]$~\cite{Leaver85,Ansorg2016,Macedo2018c,Gajic:2019qdd,Gajic:2019oem}. As in the case of static black-hole spacetimes, Leaver's approach~\cite{Leaver85} follows from a particular choice of the hyperboloidal formulation, the so-called minimal gauge --- see sec.~\ref{sec:MG_CauchyFix}.

\subsubsection{Hyperboloidal transformation in the frequency domain\\}
\label{sec:HypFreqDom}
Finally, we can compare the relation between the regular functions ${\rm v}(\sigma)$ and ${\mathbb S}(\theta)$ obtained via the Fourier transform of the hyperboloidal wave equation against the original ${\cal R}(r)$ and $S(\theta)$ introduced in~\eqref{eq:ansatzFreq_BL}. To this end, we recall the complete mapping from Boyer-Lindqst to the hyperboloidal coordinates as
\bea
\label{eq:BL_Hyp}
t = \lambda\bigg(\tau -h(\sigma, \theta) \bigg) - r^*( r(\sigma) ), \quad r=\lambda\dfrac{\rho(\sigma)}{\sigma}, \quad \varphi = \phi - k(r(\sigma)).
\eea
After inserting eq.~\eqref{eq:BL_Hyp} directly into eq.~\eqref{eq:ansatzFreq_BL}, one reads the following relation between the hyperboloidal frequency parameter $s$ and the frequency $\omega$
\beq
s = - i \lambda \omega.
\eeq
Then, the re-scaling provided by eqs.~\eqref{eq:RegField} and \eqref{eq:ReScaleUm_to_Vm} leads to
\bea
S(\theta) &=& \Xi(\theta) {\mathbb S}(\theta), \nn \\
 \Xi(\theta) &=& \cos^{\delta_1}(\theta/2) \sin^{\delta_2}(\theta/2) \exp\left[ s H_1(\theta)\right] \label{eq:FactorTheta},
 \eea
 together with
 \bea
{\cal R}(r) &=& Z(\sigma(r) )\, {\rm v}(\sigma(r)), \nn \\ 
Z(\sigma) &=& \Omega(\sigma)^{1+2p}\tilde\Delta(\sigma)^{-p} \exp\left[  s\left( H_0(\sigma) + \frac{r^*(r(\sigma))}{\lambda} \right) + im\,k(r(\sigma))\right]. \label{eq:FactorZ}
\eea
With the above consideration, one verifies the following relation between the differential operators in the original formulation and the hyperboloidal setting
\beq
\boldsymbol{ {\cal T}} S = \Xi \, \boldsymbol{\Theta} {\mathbb S}, \quad \boldsymbol{{\cal D}} \, {\cal R} =  Z\, \boldsymbol{A} {\rm v}.
\eeq
The functions $\Xi(\theta)$ and $Z(\sigma)$ are precisely the regularisation factors needed to incorporate the regularity/boundary conditions into the original eqs.~\eqref{eq:SpinWeightSphHarm} and \eqref{eq:QNM_BL}. Typically, they are determined via standard techniques in the theory of ordinary differential equations, for instance, as in the well-known works by Leaver~\cite{Leaver85,Leaver90}. Clearly, they are not unique. In fact, other choices have been considered when studying the original eq. \eqref{eq:QNM_BL} in the frequency domain, such as the ones by Dolan and Ottewill~\cite{Dolan:2009nk,Dolan:2010wr}. 

Here, we observe that the factors \eqref{eq:FactorTheta} and \eqref{eq:FactorZ} are directly related to the choice of the height function and the radial compactification in the hyperboloidal formulation. Therefore, they can be derived directly from a spacetime perspective, adding in this way an extra geometrical insight into the problem.

\medskip
The next sections discuss several hyperboloidal gauges for the Kerr spacetime and the TE, which amounts to different choices of the radial function $\rho(\sigma)$ --- or equivalently $\beta(\sigma)$  --- and the time function $A(\sigma,\theta)$.

\section{The minimal gauge}\label{sec:MG}
This section introduces the minimal gauge (MG) for the Kerr solution. As in~\cite{Macedo2018}, one requires the gauge functions $\beta(\sigma)$ and $A(\sigma,\theta)$ --- cf.~eqs.~\eqref{eq:beta} and \eqref{eq:Height} --- to assume their simplest  form
\beq
\label{eq:MinGaug}
A(\sigma,\theta) = 0, \quad \beta(\sigma) = \rho_0 \Longrightarrow \rho(\sigma) = \rho_0 + \sigma \rho_1.
\eeq
All the degrees of freedom now reduces to a choice of a preferred length scale $\lambda$, together with the values $\rho_0$ and $\rho_1$, restricted to $\rho(\sigma)>0$ --- see discussion around eq.~\eqref{eq:AsympExp_rho_beta}. 

We begin by choosing $\lambda=r_+$, which according to~\eqref{eq:KerrParm} and \eqref{eq:MassSpinPar} fixes the dimensionless Kerr parameters to
\beq
\mu = \frac{1+\kappa^2}{2}, \quad \alpha = \kappa.
\eeq
Moreover, it is convenient to set the coordinate location of the black-hole horizon $r_+$ at $\sigma_{\rm h}=1$, which constraints $\rho_0$ and $\rho_1$ to 
\bea
\rho_0&=&r_+/\lambda-\rho_1 \nn \\ &=& 1-\rho_1. 
\eea
The minimal gauge leads naturally to two distinct conformal representation of the spacetime, depending on the choice of the remaining gauge parameter $\rho_1$. The two geometries have different spacetime limits in the extremal case.
\subsection{Radial function fixing gauge} 
The most straight forward choice is to set \beq \rho_1 = 0. \eeq 
Such an option fixes the radial function to the value
\beq
\rho(\sigma)  = 1,
\eeq
and thus it is referred to as the radial function fixing minimal gauge (MG$_{\rm R}$).

It leads to a conformal line element with metric components --- cf.~eqs.~\eqref{eq:ConfMetric}-\eqref{eq:ConfDet}.
\bea
 \tilde{g}_{00} &=& -\sigma^2 F, \,\,   \tilde{g}_{01} = -\bigg(1-2F\left[ 1 + (1+\kappa^2)\sigma\right]\bigg),\,\, \tilde{g}_{22} = \tilde\Sigma,  \nn \\
 \tilde{g}_{03} &=& -\dfrac{\kappa (1+\kappa^2) \sigma^3}{\tilde\Sigma} \sin^2\theta, \,\,   \,\,
  \tilde{g}_{33} = \dfrac{\sin^2\theta}{\tilde\Sigma}\left( \tilde\Sigma_0^2 - \tilde\Delta \kappa^2\sigma^2\sin^2\theta\right), \nn \\
 \tilde{g}_{11}&=& 4\dfrac{(1+\kappa^2)}{\tilde\Sigma}\left[ 1 + (1+\kappa^2)\sigma\right]\left[1+\kappa^2(1-\sigma\cos^2\theta)\right],  \label{eq:ConfMetric_RadFuncGauge} \\ 
 \tilde{g}_{13} &=& \kappa \sin^2\theta \left(1 +  2\dfrac{(1+\kappa^2)\sigma}{\tilde\Sigma} \left[ 1 + (1+\kappa^2)\sigma\right] \right); \nn
\eea
\bea
\tilde{g}^{00} &=& -\dfrac{1}{\tilde\Sigma} \bigg( 4(1+\kappa^2) \left[ 1 + (1+\kappa^2)\sigma\right] \left[   1 +  \kappa^2(1-\sigma)  \right] - \kappa^2\sin^2\theta \bigg)   \nn \\
\tilde{g}^{11} &=& \dfrac{\sigma^2\tilde\Delta}{ \tilde\Sigma},\quad \tilde{g}^{22} = \dfrac{1}{\tilde\Sigma}, \quad \tilde{g}^{33} = \dfrac{1}{\tilde\Sigma\sin^{2}\theta}, \quad \tilde{g}^{13} = -\dfrac{\kappa \sigma^2 }{\tilde\Sigma}, \label{eq:ConfInvMetric_RadFixGauge}  \\
\tilde{g}^{01} &=& -\dfrac{1}{\tilde\Sigma}\left(\tilde\Sigma_0 - 2 \tilde\Delta \left[ 1 + (1+\kappa^2)\sigma\right] \right), \, \tilde{g}^{03} = \dfrac{\kappa}{\tilde\Sigma}\bigg(1 - 2\left[ 1 + (1+\kappa^2)\sigma\right]\bigg); \nn
\eea
\bea
\label{eq:ConfDet_RadFuncGauge}
\tilde g &=& \det\tilde g_{ab} =  - \tilde\Sigma^2 \, \sin^2\theta.
\eea
The Kerr spacetime in the MG$_{\rm R}$ gauge is well-defined in the complete parameter range $\kappa\in[-1,1].$ In particular, the limiting value $|\kappa|=1$ approaches the usual extremal Kerr spacetime. Indeed, the norm of the vector $\tilde\nabla_a\Omega$ at $\sigma=0$ indicates the type of hypersurface $\scri^+$ is. In this gauge, we obtain
\beq
\lim_{\sigma \rightarrow 0 } |\tilde\nabla_a \Omega|^2 = \lim_{\sigma \rightarrow 0 } \dfrac{\sigma^2(1-\sigma)(1-\kappa^2\sigma)}{r_+^2 \tilde\Sigma} = 0, \quad \forall \, \kappa\in[-1,1].
\eeq 
In other words, $\scri^+$ is a null hypersurface for all values of the parameter $\kappa$, including the extremal case $|\kappa|=1$.

The Cauchy horizon $r_-$ in the new compact radial coordinate is $\sigma_- = \kappa^{-2}$. Thus, the coordinate location $\sigma_-$ changes in the $\sigma$-direction parametrically, according to $\kappa$. In particular, at the Schwarzschild limit $\kappa = 0$, $\sigma_-\rightarrow \infty$ , i.e.~it corresponds to the singularly $r_- =0$. In the extremal case $\kappa=1$, the horizons coincide at $\sigma_- = \sigma_+ = 1$, as expected.

\ref{App:TE_MG} displays the complete form of the Teukolsky equation in the time and frequency domain for the MG$_{\rm R}$ gauge. Here, we focus on the factor $Z(\sigma)$ that regularises the TE in the frequency domain according to eq.~\eqref{eq:FactorZ}
\bea
Z(\sigma) &\propto& \left(1-\dfrac{r_+}{r(\sigma)}\right)^{-p+\frac{im\kappa + 2\mu s}{1-\kappa^2}} \left(1-\kappa^2\dfrac{ r_+}{r(\sigma)}\right)^{-p-\frac{im\kappa + 2\mu s}{1-\kappa^2}} \nn \\
&&\times \exp\left( -s\dfrac{r(\sigma)}{r_+}\right) \left(\dfrac{r(\sigma)}{r_+} \right)^{-(1+2p+2\mu s)}. \label{eq:FacZ_MGR}
\eea
Of particular interest is the expression of eq~\eqref{eq:FacZ_MGR} in the extremal limit $| \kappa| \rightarrow 1$ ($r_+ = M$)
\bea
Z(\sigma) &\propto&  \exp\left[ -s\dfrac{r(\sigma)}{M} - (2s + im)\left( \dfrac{r(\sigma)}{M} -1 \right)^{-1}\right]  \nn \\
&&\times \left(1-\dfrac{M}{r(\sigma)}\right)^{-2p +2 s}\left(\dfrac{r(\sigma)}{r_+} \right)^{-(1+2p+2s)}. \label{eq:FacZ_MGR_Ext}
\eea
Eq.~\eqref{eq:FacZ_MGR_Ext} corresponds precisely to the factor introduced by Richartz in ref.~\cite{Richartz:2015saa} to calculate the quasi-normal modes of an extremal Kerr black hole according to Leaver's algorithm. Within the hyperboloidal approach, it becomes clear that spacetime counterpart of Richartz ansatz is the description of the Kerr solution in the radial function fixing minimal gauge.

\subsection{Cauchy horizon fixing gauge} 
\label{sec:MG_CauchyFix}
As an alternative to the previous gauge, one can fix the Cauchy horizon at a pre-defined coordinate distance $\sigma_-=c^{-1}$, {\em independent} of the Kerr parameter $\kappa$. This family of gauges is called the Cauchy horizon fixing minimal gauge (MG$_{\rm C}$).

One natural requirement is that both the MG$_{\rm R}$ and the MG$_{\rm C}$ gauges lead to the same results in the Schwarzschild limit $\kappa=0$. Such a requirement gives $c=0$, i.e., it fixes the Cauchy horizon at $\sigma_- \rightarrow \infty,$ $\forall \kappa$.
This property is achieved by setting \beq\rho_1 = \kappa^2,\eeq which leads to the metric components
\bea
 \tilde{g}_{00} &=& -\sigma^2 F, \,\, \tilde{g}_{22} = \tilde\Sigma, \,\, \tilde{g}_{33} = \dfrac{\sin^2\theta}{\tilde\Sigma}\left( \tilde\Sigma_0^2 - \tilde\Delta \kappa^2\sigma^2\sin^2\theta\right), \nn  \\   
 \tilde{g}_{11}&=& \dfrac{4}{\tilde\Sigma}(1+\kappa^2) [1-\kappa^2 + (1+\kappa^2)\sigma]\left[ 1-\kappa^2(1-\sigma \sin^2\theta)  \right], \nn \\ 
 \tilde{g}_{01} &=& - (1-\kappa^2)(1-2F) + 2F (1+\kappa^2)\sigma , \label{eq:ConfMetric_CauchyFuncGauge} \\
  \tilde{g}_{03} &=& -  \dfrac{\kappa(1+\kappa^2) [1-\kappa^2(1-\sigma)] \sigma^3}{\tilde\Sigma} \sin^2\theta. \nn \\
 \tilde{g}_{13} &=& \kappa \sin^2\theta \left(1-\kappa^2 +  2\sigma\dfrac{(1+\kappa^2) [1-\kappa^2(1-\sigma)][1+\sigma - \kappa^2( 1-\sigma)]}{\tilde\Sigma}  \right) \nn
\eea
\bea
\tilde{g}^{00} &=& -\dfrac{1}{\tilde\Sigma} \Bigg( 4(1+\kappa^2) \left[ 1 + \dfrac{1+\kappa^2}{1-\kappa^2}\sigma\right]  - \kappa^2\sin^2\theta \Bigg) , \nn \\ 
\tilde{g}^{11} &=& \dfrac{\sigma^2 (1-\sigma)}{ \tilde\Sigma}, \, \tilde{g}^{22} = \dfrac{1}{\tilde\Sigma}, \, \tilde{g}^{33} = \dfrac{1}{\tilde\Sigma\sin^{2}\theta}, \, \tilde{g}^{13} = -\dfrac{\kappa \sigma^2 }{(1-\kappa^2)\tilde\Sigma} \nn \\
\tilde{g}^{01} &=& -\dfrac{1}{(1-\kappa^2)\tilde\Sigma} \Bigg( \tilde\Sigma_0- 2\tilde\Delta\left[ 1 + \dfrac{ (1+\kappa^2)\sigma}{1-\kappa^2}\right] \Bigg), \nn \\
\tilde{g}^{03} &=& \dfrac{\kappa}{\tilde\Sigma}\left(1 - 2\left[1 +\dfrac{1+\kappa^2}{1-\kappa^2}\sigma\right]\right); \nn
\eea
\bea
\label{eq:ConfDet_CauchyFuncGauge}
\tilde g &=& \det\tilde g_{ab} =  -(1-\kappa^2)^2 \, \tilde\Sigma^2 \, \sin^2\theta.
\eea
By writing the Teukolsky equation in the Cauchy horizon fixing gauge --- see \ref{App:TE_MG} --- one observes that the MG$_{\rm C}$ gauge provides the same regularisation scheme as Leaver's approach to the TE in the frequency domain~\cite{Leaver85}. Indeed, one first notices that, in terms of the original radial coordinate $r(\sigma)$, the Taylor expansion around the horizon for ${\rm v}(\sigma)$ in eq.~\eqref{eq:Taylor_v} reads
\beq
{\rm v}(\sigma) = \sum_{n=0}^{\infty} {\rm b}_n \left(1-\sigma\right)^n = \sum_{n=0}^{\infty} {\rm b}_n \left(\dfrac{r(\sigma)-r_+}{r(\sigma)-r_-}\right)^n,
\eeq
i.e., it leads precisely to the expansion used by Leaver. Apart from that, the hyperboloidal regularisation factor \eqref{eq:FactorZ} reads
\bea
Z(\sigma) &\propto& \exp\left(-sr(\sigma)/r_+\right)\left(\dfrac{r(\sigma)}{r_+}\right)^{-(1+2p)}  \left( 1-\dfrac{r_+}{r(\sigma)} \right)^{-p+(2s\mu+im\kappa)/(1-\kappa^2)}\times \nn \\
 &&\times
\left(1-\dfrac{r_-}{r(\sigma)}\right)^{-(1+p)-[2\mu s(2-\kappa^2) im\kappa]/(1-\kappa^2)},
\eea
which is precisely the one used by ref.~\cite{Leaver85}.

The strategy employed by Leaver for regularising the TE in the frequency domain is based on tools developed for the study of ordinary differential equations. Here, we see that the MG$_{\rm C}$ gauge provides a complete spacetime description of Leaver's approach.

\medskip
A well-known limitation in Leaver's strategy is that it does not apply to the extremal case $|\kappa|= 1$. By having the spacetime description, it becomes evident that the extremal limit $|\kappa|\rightarrow 1$ is not well defined due to the vanishing of the determinant $\tilde{g}|_{\kappa=1} =0$ --- or equivalently, by the singular behaviour of some components for the inverse metric $\tilde{g}^{ab}$. Such a singular behaviour reflects the fact that the radial transformation~\eqref{eq:Coords_Hyp} is ill-defined in the extremal limit, i.e., $r(\sigma) \stackrel{|\kappa| \rightarrow 1}\longrightarrow 1$. 

These results on the limit to extremality are not restricted to the choice of fixing the Cauchy horizon at $\sigma_- \rightarrow \infty,$ $\forall \kappa$. It is valid whenever one fixes the Cauchy horizon to a given coordinate location $\sigma_-$ independent of the parameter $\kappa$.

Interestingly, one observes that $\scri^+$ becomes a timelike surface in the extremal limit. Indeed, one has
\bea
\lim_{\sigma \rightarrow 0 } |\tilde\nabla_a \Omega|^2 &=& \lim_{\sigma \rightarrow 0 } \dfrac{\sigma^2 (1-\sigma)}{r_+^2 (1-\kappa^2)(1-\kappa^2 -2\sigma) + \kappa^2\sigma^2(\kappa^2+\cos^2\theta)} \nn \\
&=& \left\{
\begin{array}{cc}
0 & \kappa\in(-1,1) \\
r_+^{-2}(1+\cos^2\theta)^{-1} > 0 & |\kappa|=1
\end{array}
\right. .
\eea
The timelike character of the spacetime boundary is typical of (asymptotically) Anti-de Sitter manifolds, and therefore it strongly suggests that the limit corresponds actually to the near-horizon geometry~\cite{Bardeen:1999px,Kunduri2013}. 

The near-horizon geometry is achieved by one further coordinate transformation
\beq
\tau = \dfrac{T}{1-\kappa^2}, \quad \phi = \Phi + \dfrac{\kappa}{(1+\kappa^2)(1-\kappa^2)}T - 2\ln\sigma.
\eeq
As in the radial case, the transformation above is ill-defined for $\kappa\rightarrow 1$. Nevertheless, the resulting metric is regular in the limit. In particular, the $T$ dependence in the angular coordinate is crucial to ensure the regularity of the final line element. On the other hand, the $\sigma$ dependence is not needed for obtaining a regular limit $\kappa \rightarrow 1$. However, it simplifies the final result as it implies that $T$ is an ingoing Eddington-Finkelstein-type null coordinate. In the final coordinates $(T,\sigma,\theta,\Phi)$, the conformal line element for $\kappa=1$ reads
\bea
\d \tilde{s}^2 &=& (1+\cos^2\theta)\left( -\dfrac{1-\sigma}{4}\d T^2 - \d T \d \sigma + \sigma^2\d\theta^2\right) \nn \\ 
&& + \dfrac{4\sin^2\theta}{1+\cos^2\theta}\left( \sigma \d \Phi + \dfrac{1-\sigma}{2} \d T\right)^2.
\eea
Therefore, the failure of Leaver's algorithm in the limit $|\kappa|\rightarrow1$ is not a technical one. It is instead a consequence of the discontinuous transition to the near-horizon geometry. 

\medskip
The next section surveys several hyperboloidal gauges in the literature and presents them in terms of a single formalism for comparison. It becomes evident the advantages of the minimal gauge due to its simplicity.

\section{Further hyperboloidal gauges}\label{sec:HypGauges}
We recall that our compact radial coordinate $\sigma$ is naturally adapted to the conformal factor $\Omega$ via eq.~\eqref{eq:ConfFact}, and the radial gauge degree of freedom is incorporated by the function $\rho(\sigma)$. Most works, however, followed Zenginoglu's scri fixing approach~\cite{Zenginoglu:2007jw}, and employed the following radial compactification
\beq
r = \dfrac{R}{\Omega(R) } = \lambda \dfrac{R}{\sigma(R)}.
\eeq
In other words, the gauge freedom is encoded in the conformal factor $\Omega(R)=\sigma(R)/\lambda$. The re-construction of the radial function $\rho(\sigma)$ follows straightforwardly from $\rho(\sigma) = R(\sigma)$.

The motivation in~\cite{Zenginoglu:2007jw} to let the conformal factor $\Omega(R)$ free comes from the community's initial objective of applying the hyperboloidal approach to solve numerically the full non-linear Einstein's equation\footnote{See \cite{Hubner:1999th,Frauendiener2002,Bardeen:2011ip,Rinne:2009qx,Rinne:2013qc,Vano-Vinuales:2014koa,Morales:2016rgt,Hilditch:2016xzh,Vano-Vinuales:2017qij,Gasperin:2018tvk,Gasperin:2019rjg} for studies focusing on several numerical aspects of non-linear time evolutions in the hyperboloidal approach.}. Since the background spacetime is known a priori in black-hole perturbation theory, we argue here in favour of the compact coordinate adapted to the conformal factor $\sigma=\lambda\Omega$. This choice simplifies the equations involved significantly, especially in the minimal gauge discussed in the previous section.

One must also pay careful attention to the definition of the height function when comparing the different gauges available in the literature. Here, the height function $h(\sigma,\theta)$ follows from an {\em advanced} time coordinate $v$ --- cf.~eqs. \eqref{eq:TrasnfBL_KS} and \eqref{eq:Coords_Hyp} --- whereas Zenginoglu introduces his height function $h_{\rm Z}(r)$ out of a ``standard" time coordinate\footnote{More precisely, along the ``standard" time surface $t=$ constant, the limit $r\rightarrow \infty$ leads to spacelike infintiy $i^0$. } $t$ and with the opposite sign~\cite{Zenginoglu:2007jw}. Since, the time coordinates relate via $v\sim t + r$, or more precisely $v = t + r^*$, one usually obtains $h = -(h_{\rm Z} + r/\lambda)$ or $h = -(h_{\rm Z} + r^*/\lambda)$ depending on the application.

Finally, we noticed that several works bring expressions mixing dimensionful and dimensionless quantities. Their approach is justified as one eventually sets $M=1$ in the numerical experiments. However, before this scaling effectively takes place, the interpretation of such expressions adds another layer of difficulty when comparing the formalisms. Here, the hyperboloidal coordinates $(\tau, \sigma, \theta, \varphi)$ are dimensionless, and so are the functions $h$ and $\rho$. The conformal factor, on the other hand, has dimension $[\Omega] = ({\rm Length })^{-1}$. Hence, we consistently keep track of the generic length scale $\lambda$ to ease the dimensional analysis of all final expressions. Thus, setting $M=1$ is equivalent to the choice $\lambda=M$ --- see \ref{App:LengthScale}. 

\subsection{Zenginoglu's gauge \\}
In~\cite{Zenginoglu:2007jw}, Zenginoglu fixes the conformal factor to $\Omega = \lambda^{-1} (1-R) \Rightarrow \sigma = 1-R$. Therefore, one reads the radial function directly
\beq
\rho_{\rm Z}(\sigma) = 1 - \sigma.
\eeq
The horizons are located at the coordinate value $\sigma_{\pm} = (1+r_{\pm}\lambda^{-1})^{-1} <1$.

Then, he constructs two hyperboloidal coordinates for the Kerr spacetime:
\ben
\item His asymptotic regularisation of the Kerr metric initially in the Boyer-Lindquist coordinates led to $h_{\rm Z_{BL}}(r) = \dfrac{r}{\lambda} + 2\mu \ln\left(\dfrac{r}{\lambda}\right)$. Since $v = t_{\rm BL} + r^*(r)$, one obtains:
\bea
 A_{\rm Z_{BL}}(\sigma) &=& - \Bigg( h_0(\sigma) + h_{\rm Z_{BL}}( r(\sigma) ) +\dfrac{r^*(r(\sigma))}{\lambda} \Bigg) \nn \\
&=& 2  + 2 \mu \ln\left(\dfrac{r_+}{\lambda}\right) - 2\mu\ln(1-\sigma) \nn  \\
&& - \dfrac{2\mu}{1-\kappa^2}\left[ \ln\left(1-\dfrac{\sigma}{\sigma_+}\right)-\kappa^2 \ln\left(1-\dfrac{\sigma}{\sigma_-}\right) \right]  . 
\eea
Note that this choice leads to a hyperboloidal foliation which is not horizon penetrating.

\item His asymptotic regularisation of the Kerr metric initially in the Kerr-Schild coordinates led to $h_{\rm Z_{KS}}(r) = \dfrac{r}{\lambda} + 4\mu \ln\left(\dfrac{r}{\lambda}\right)$. Since $v = t_{\rm KS} + r$, one obtains
\bea
&& A_{\rm Z_{KS}}(\sigma) = - \Bigg( h_0(\sigma) + h_{\rm Z_{KS}}( r(\sigma) ) +\dfrac{r(\sigma)}{\lambda} \Bigg) \nn \\
&& = 2\Bigg( 1 - 2\mu \ln(1-\sigma)\Bigg).
\eea
\een

\subsection{The RT gauge\\}
Following Moncrief's~\cite{Moncrief2000} construction of hyperbolas in the Minkowski spacetime, R\'acz and T\'oth~\cite{Racz:2011qu} added the logarithm contribution to the height function, needed in any black-hole spacetime. 

In their gauge, one reads the conformal factor $\Omega = \lambda^{-1}(1-R^2)/2\rightarrow \sigma = (1-R^2)/2$, from which the radial function becomes
\beq
\label{eq:rho_HH}
\rho_{\rm RT}(\sigma) = \sqrt{1-2\sigma}.
\eeq
The coordinate location of the horizons are $\sigma_{\pm} = \left(\lambda r_\pm^{-1}\right)^2\bigg(\sqrt{1+r_{\pm}\lambda^{-1}}-1\bigg)$.

Their height function reads $h_{\rm RT}(r) = \sqrt{1+ (r\lambda^{-1})^2} -4\mu\ln\bigg(2\sigma(r) \bigg)$, and it was introduced according to Zenginoglu's formalism into the Kerr metric originally written in the Kerr-Schild form. Since $v= t_{\rm KS} + r$, one gets the function
\bea
&& A_{\rm RT}(\sigma) = - \Bigg( h_0(\sigma) + h_{\rm RT}( r(\sigma) ) +\dfrac{r(\sigma)}{\lambda} \Bigg) \nn \\
&& = 1+ 4\mu\ln 2 + \dfrac{1-\sqrt{1-2\sigma}}{\sigma}
\eea 
It is easy to see that at $\sigma=0$ is, indeed, regular.

This gauge was then used by Harms and Bernuzzi as well~\cite{Harms:2013ib}. 
\subsubsection{The HH$_S$ gauge \\}
Harms, Bernuzzi, Nagar and Zenginoglu introduced the HH$_S$ with a free parameter $S$ adapting their numerical scheme~\cite{Harms:2014dqa}. In the HH$_S$ gauge, one reads the conformal factor $\Omega = \lambda^{-1}(1-R/S)\rightarrow \sigma = (1-R/S)$, from which the radial function becomes
\beq
\label{eq:rho_HHS}
\rho_{\rm HH_S}(\sigma) = S(1-\sigma).
\eeq
The coordinate location of the horizons are $\sigma_{\pm} =\left(1+\dfrac{r_{\pm}}{\lambda S}\right)^{-1} <1$.

For their numerical studies, they demand invariance of the coordinate expression for outgoing characteristics in the spatially compactified coordinates. Hence, they express the advanced-time coordinate in the Kerr-Schild form as \beq v/\lambda=\tau + \rho(\sigma)\dfrac{S+\rho(\sigma)}{S-\rho(\sigma)} - 4\mu \ln( \sigma) - 2\mu \ln(2\mu), \nn \eeq
from which we read 
\beq
\label{eq:A_HH}
A_{\rm HH_S}(\sigma) = 2\mu\ln\mu +3S - S\sigma.
\eeq

\medskip
In the author's original coordinates $R$, the parameter $S$ is interpreted as the (free) coordinate location of $\scri^+$. Here though, where one always has $\sigma_{\scri^+} = 0$, the gauge HH$_S$ is viewed as the simplest  extension away from the minimal gauge. Indeed, as required by the minimal gauge, eq.~\eqref{eq:rho_HH} leads to $\beta_{\rm HH_S}=S=$ constant, whereas eq.~\eqref{eq:A_HH} introduces only\footnote{In all gauges, constants within $A(\sigma)$ affect only an overall time-offset.} a linear term $A_{\rm HH_S}(\sigma) \sim -S\sigma $.

Despite the simple modification, the current formulation of the HH$_S$ gauge does not provide a smooth transition to the minimal gauge as $S\rightarrow 0$ because the radial transformation is not defined for $S=0$ --- cf.~\eqref{eq:rho_HHS}. Therefore, this work introduces the mHH$_S$, i.e., a modified, but equivalent version of the HH$_S$ gauge by adopting simply
\beq
\rho_{\rm mHH_S}(\sigma) = 1+S(1-\sigma), \quad A_{\rm mHH_S}(\sigma) = -S\sigma. \label{eq:mHHS}
\eeq
With this choice, one has $\scri^+$ and ${\cal H}^+$ respectively fixed at $\sigma=0$ and $\sigma=1$, regardless of $S$. Moreover, $S\rightarrow 0$ naturally recovers the MG$_{\rm R}$ gauge.

\subsection{Tha Dolan and Ottewill gauge in the frequency domain}
All hyperboloidal gauges for the Kerr spacetime available in the literature treat the problem in the time domain. With a focus on the frequency domain, this work has already identified the MG$_{\rm C}$ gauge as the counterpart of Leaver's~\cite{Leaver85,Leaver90} approach. 

This section discusses from the hyperboloidal perspective the approach by Dolan and Ottewill~\cite{Dolan:2009nk,Dolan:2010wr} to approximate the quasi-normal spectrum of black holes in the Eikonal limit. In particular, ref.~\cite{Dolan:2010wr} introduces the Ansatz\footnote{The notation in this section is kept as close as possible to the work\cite{Dolan:2010wr}. Hence, the functions $\beta_{\rm D}$ and $\Omega_{\rm D}$ must not be confused with the radial shift $\beta$, nor with the conformal factor $\Omega$ introduced previously in this work.}
\beq
\label{eq:DolanAnsatz}
R(r) = r^{-1} \Delta \exp\left(i \int\beta_{\rm D}(r) d\bar{r} \right) {\rm v}(r), \quad \dfrac{d\bar r}{dr} = \dfrac{r^2}{\Delta}
\eeq 
into the Teukolsky  radial equation \eqref{eq:QNM_BL}.

Then, ref.~\cite{Dolan:2010wr} demands that $\beta_{\rm D}$ satisfies: (i) the appropriate outgoing/ingoing boundary conditions at infinity/horizon; (ii) the resulting differential equation for ${\rm v}$ allows the factorisation of an overall term $\Delta/r^2$; and (iii) $\beta_{\rm D}$ changes sign when crossing a particular $r_{\rm orb} $ corresponding to unstable circular orbits of null geodesics. 

Since there is no mention to a radial compactification in their work, we assume for simplicity the simplest  relation $r={\lambda}/{\sigma}.$, i.e., with the radial function
\beq
\rho_{\rm DO} = 1.
\eeq
A comparison between the Ansatz \eqref{eq:DolanAnsatz} against the hyperboloidal expression \eqref{eq:FactorZ} leads to
\bea
\beta_{\rm D} &=& \beta^{(0)}_{\rm D} + \omega \sigma^2 \tilde\Delta \dfrac{\d A_{\rm DO}}{\d \sigma}, \quad \beta^{(0)}_{\rm D} = \Omega_{\rm D} + 2\omega \tilde\Delta (1+ 2\mu\sigma), \nn \\ \Omega_{\rm D} &=& (1+\alpha^2\sigma^2)\omega + m\alpha\sigma^2/\lambda.
\eea 
Thanks to the hyperboloidal construction, $\beta_{\rm D}$ satisfies conditions (i) and (ii) automatically. In terms of the hyperboloidal function $A(\sigma)$, one gets for the Dolan gauges
\beq
\label{eq:FuncADolan}
\dfrac{\d A_{\rm DO}}{\d \sigma} = \dfrac{1}{\sigma^2} \left( - 2\left(1+2\mu \right) + \dfrac{1}{\omega}\left[\dfrac{\beta_{\rm D}+\Omega_{\rm D}}{\tilde \Delta}\right] \right).
\eeq
As for condition (iii), the ref.~\cite{Dolan:2010wr} considers equatorial and polar orbits. 
\subsubsection{Equatorial orbit \\}
For the unstable null geodesics in the equatorial plane, the function $\beta_{\rm D}$ reads
\beq
\label{eq:betaDolan_EqOrb}
\beta_{\rm D_{eo}}=\Omega_{\rm D}\left( 1 - \alpha(b_{\rm eo} - \alpha)\sigma^2 \right)^{-1} \left( 1 - \hat{r}_{\rm eo}\sigma\right)\left( 1 + s\hat{r}_{\rm eo}\sigma\right)^{1/2},
\eeq
with $\hat{r}_{\rm eo} = 2\mu\left[ 1+\cos(2 \arccos(-\alpha/\mu) /3)\right]$ the dimensionless orbit radius and $b_{\rm eo} = 3\sqrt{\mu \hat{r}_{\rm eo}} - \alpha$ the dimensionless impact parameter. By expanding eq.~\eqref{eq:FuncADolan} and \eqref{eq:betaDolan_EqOrb} around $\sigma=0$ one verifies that the resulting $A_{\rm DO_{eo}}$ is regular at $\scri^+$.

Note that $A_{\rm DO_{eo}}$ retains a parameter $m/\omega$. When working in the frequency domain, the algorithm developed in~\cite{Dolan:2010wr} eventually associates the frequency $\omega$ the black-hole's quasi-normal modes. The usage of this gauge in the time domain would require a re-interpretation of the frequency since the function $A(\sigma)$ should be real-valued. One possible choice is to work with the energy of the photon in the circular orbit.  Further studies interpreting and discussing the advantages of this gauge in the time domain are required, and they go beyond the scope of this work.

\subsubsection{Polar orbit \\}
For the unstable null geodesics in polar orbits, the function $\beta_{\rm D}$ reads
\beq
\label{eq:betaDolan_PolarOrb}
\beta_{\rm D_{po}}=\omega \left( 1-\hat r_{\rm po} \sigma \right)\sqrt{1+2\hat r_{\rm po}\sigma - \dfrac{\alpha^2(b_{\rm po}^2-\alpha^2)}{\hat r_{\rm po}}\sigma^2} 
\eeq
with $\hat{r}_{\rm po} = \mu + 2\sqrt{\mu^2 - \alpha^2/3} \cos\left[3^{-1} \arccos\left( \dfrac{\mu(\mu^2-\alpha^2)}{(\mu^2-\alpha^2/3)^{3/2}}\right) \right]$ the dimensionless orbit radius and $b_{\rm po} = \sqrt{\dfrac{(3\hat{r}_{\rm po}^2-\alpha^2)(\hat{r}_{\rm po}^2+\alpha^2)}{\hat{r}_{\rm po}^2 - \alpha^2}}$ the dimensionless impact parameter. As in the previous case,  an expansion around $\sigma=0$ shows that resulting $A_{\rm DO_{po}}$ is regular at $\scri^+$.

Contrary to the orbits in the equatorial plane, polar orbits have $m=0$. Therefore, factors containing the frequencies $\omega$ cancels out and this gauge is suitable for eventual studies and evolutions directly in the time domain.

\subsection{The $\theta$-dependence}
So far, the functions $A$ considered have only a radial dependence. We end this section by mentioning further gauges in the literature where an angular dependence is present.

\subsubsection{The CMC and ACMC gauges \\}
Of great interest in the study of the conformal Einstein's equations are the so-called constant mean curvature (CMC) slices. Contrary to the Schwarzschild spacetime, in which CMC slices are known analytically~\cite{Malec:2009hg}, in the Kerr spacetime, they are only obtained numerically~\cite{Schinkel:2013tka}. For their construction one: (i) writes down the mean curvature $K(\sigma,\theta)$ in terms of the function $A(\sigma,\theta)$, and (ii) solves the resulting second-order differential equation for the unknown $A(\sigma,\theta)$ which results from the condition $K(\sigma,\theta)=K_0$ constant. In~\cite{Schinkel:2013tka}, we found regular solutions for all parameters $\kappa\in[0,1]$.

By relaxing the global CMC condition~\cite{Schinkel:2013zm}, we introduced analytic --- though lengthy ---  functions $A(\sigma,\theta)$ leading to hypersurfaces in which the mean curvature behaves as $K(\sigma,\theta)= K_0 + {\cal O}(\sigma^4)$, the so-called asymptotically constant mean curvature  (ACMC) condition.

\subsubsection{The Newman-Janis ``complex gauge" \\}
\label{sec:NewmanJanis_Leaver}

Finally, it is interesting to notice that the Newman-Janis complexification's algorithm~\cite{NewmanJanis} to derive the Kerr metric from the Schwarzschild solution introduces a complex transformation of the null coordinate $v$. This ``trick" can be formally absorbed in our framework by
\beq
\label{eq:ComplexA}
A(\sigma,\theta) = i \alpha\cos\theta.
\eeq
At this stage, the above expression is regarded just as a side remark in the formalism. Indeed, if one were to take eq.~\eqref{eq:ComplexA} into the definition of the hyperboloidal coordinates~\eqref{eq:Coords_Hyp} and \eqref{eq:Height}, one would obtain a complex time $\tau$. Nevertheless, when considering the Teukolsky equation in the frequency domain, eq.~\eqref{eq:ComplexA} leads precisely to the normalisation factor introduced by Leaver~\cite{Leaver85} when studying the spin-weighted spheroidal harmonics --- see section~\ref{sec:TeukEq_FreqDomain}, in particular, eq.~\eqref{eq:LeaverAngHeight}.

\section{Discussion and conclusion}
In this work, we developed a comprehensive hyperboloidal framework for the Kerr spacetime. Apart from performing a systematic study of the degrees of freedom involved, the formalism provides the tools to study the Teukolsky equation in the time and frequency domains. 

The first step introduces a generic hyperboloidal coordinate system $(\tau, \sigma, \theta, \phi)$ allowing a conformal compactification of the spacetime along the spatial directions\footnote{A larger class of hyperboloidal slices can be explored by relaxing the need of the conformal compactification~\cite{Calabrese:2005rs}. In terms of the radial transformation introduced in ref.~\cite{Calabrese:2005rs}, this work has $n=2$.}. In particular, the radial coordinate $\sigma$ is naturally adapted to the conformal factor $\Omega$ via $\sigma = \lambda \Omega $, with $\lambda$ a generic length scale. The degrees of freedom are encoded by a radial function $\rho(\sigma)$ and a height function $h(\sigma, \theta)$. Then, the requirement that the resulting hypersurfaces of constant time intersect future null infinity fixes a leading term $h_0(\sigma)$ of the height function as expressed in eqs.~\eqref{eq:Height} and \eqref{eq:HeightMin}. Already observed in previous works~\cite{Zenginoglu:2007jw}, the presence of a black hole with mass $M$ introduces a logarithmic term in height function, but no further contribution from the black hole's angular momentum is required.  A similar feature was observed in~\cite{Macedo2018}, i.e., the black hole's charge does not impose any further restrictions on the leading terms of the height function. 

The systematic construction of the hyperboloidal slices should be valid for any other possible black-hole spacetime arising in modified theories of Gravity, such as \cite{Olmo:2011ja,Wei:2014dka,Chen:2018mkf}. The particular form of the height function's leading terms is a direct consequence of the asymptotic behaviour of the tortoise coordinate $dr^*/dr \sim 1 + 2M/r + {\cal O}(r^{-2})$. Thus, no matter how the modifications look like near the horizon, one should expect qualitatively the same behaviour for the height function. In this way, we anticipate that the logarithm term is a feature exclusive of $4$-dimensional black-hole spacetimes. In asymptotically flat, higher dimensions spacetimes, the leading singular contribution $h_0(\sigma)\sim \sigma^{-1}$ to the height function is sufficient for the construction of hyperboloidal slices. 

\medskip
In a second stage, we consider the Teukolsky equation, which had already been studied in the time domain for some specific hyperboloidal gauges. This work approaches the problem from a generic perspective and develops the formalism in both the time and frequency domain. Exploiting the axial-symmetry of the system, we make use of a Fourier decomposition in the azimuthal direction to write the TE as a $2+1$ evolution problem for a regular master field $V_m$. The final equation \eqref{eq:Teukosly_Vm} is to be solved in the domain $(\tau, \sigma, x)\in [\tau_0,\tau_1]\times[0,\sigma_{\rm horizon}]\times[-1,1]$ after the prescription of initial data for the field and its time derivative. As expected, no external boundary conditions are required due to the hyperboloidal nature of the times slices. Then, the hyperboloidal framework is applied to the frequency domain. Ideally, one wants to separate the equations into ordinary differential equations for the angular and radial directions. A pre-requisite for the separability of the equations in the frequency domain is that the height function also separates as $h(\sigma,\theta) = H_0(\sigma) + H_1(\theta)$. The regularisation factors for the standard Teukolsky  radial and angular equations follows straightforwardly from hyperboloidal gauge degrees of freedom --- see eqs.~\eqref{eq:FactorTheta} and \eqref{eq:FactorZ}.

Of particular importance, this work introduces the minimal gauge (MG) for the outer region of the Kerr spacetime. Its construction follows from retaining only the minimal requirement for the slices $\tau =$ constant to foliate future null infinity. For the height function, this property translates into fixing $h(\sigma,\theta) = h_0(\sigma)$, whereas the radial function $\rho(\sigma)$ reduces to a polynomial of the first order --- see eqs.~\eqref{eq:MinGaug}. Despite its simplicity, the minimal gauge provides a rich structure to study  the limits to extremality. In a first option, one can fix the radial function to a constant --- the so-called radial function fixing gauge MG$_{\rm R}$. In this gauge, the coordinate location of the Cauchy horizon $\sigma_-$ changes parametrically according to the spin parameter of the Kerr solution. As the black-hole's spin increases, the Cauchy horizon continuously approaches the event horizon, and one obtains the standard extremal Kerr black hole in the limit $|a|\rightarrow M$. A second option is to fix the Cauchy horizon at a given coordinate value independently of the spin parameter --- the so-called Cauchy horizon fixing gauge MG$_{\rm C}$. This choice leads to a discontinuous transition to Kerr's near-horizon geometry in the extremal limit~\cite{Bardeen:1999px,Kunduri2013}. The same feature was qualitatively observed in the Reisnner-Nordstrom solution~\cite{Macedo2018}.

Interestingly enough, when exploring the minimal gauge in the frequency domain, one observes that the Cauchy horizon fixing gauge corresponds exactly to Leaver's strategy to regularise the radial equation~\cite{Leaver85}. While Leaver's regularisation factor follows from techniques for ordinary differential equations, here we obtain it directly from the geometrical arguments in the hyperboloidal formalism. The spacetime insight explains the limitation in Leaver's algorithm in the extremal limit. As mentioned, the extremal limit in MG$_{\rm C}$ gauge is discontinuous, and one obtains the Kerr's near-horizon geometry. Since there is a change in topology from an asymptotically flat to an asymptotically AdS spacetime, external boundary conditions are suddenly required at $\scri^+$. Alternatively, the extremal case is treated by a Leaver-like algorithm in ref.~\cite{Richartz:2015saa}, where the regularisation is performed {\em after} the extremal limit is taken. From the hyperboloidal perspective, the strategy from ref.~\cite{Richartz:2015saa} is a consequence of writing the Kerr spacetime in the radial function fixing gauge.

A few hyperboloidal gauges have already been used to perform the time evolution of Teukolsky equation~\cite{Racz:2011qu,Jasiulek:2011ce,Harms:2013ib,Yang:2013uba,Spilhaus:2013zqa,Macedo2014,Csukas:2019kcb}. Here, we review all gauges and write them into the formalism presented in this paper. It becomes evident that the minimal gauge provides the simplest structure for (semi-)analytical studies of the Teukolsky equation. In particular, ref.~\cite{Macedo2014} studies the TE in the time domain within the MG$_{\rm R}$ gauge. Due to the implicit nature of the fully spectral code in ref.~\cite{Macedo2014}, time integration is not restricted by Courant-Friedrichs-Lewy conditions. Therefore, a direct comparison with other codes to assess the numerical efficiency of the MG via explicit time integrators requires further investigation. Several studies~\cite{Harms:2014dqa,Nagar:2014kha,Harms:2015ixa,Harms:2016ctx,Lukes-Gerakopoulos:2017vkj,Wardell:2014kea,Thornburg:2016msc} are based on the ${\rm HH_S}$ gauge, introduced initially in ref.~\cite{Harms:2014dqa}. This gauge has a free parameter $S$ to enhance the numerical stability of the code~~\cite{Harms:2014dqa}. In this paper, we showed that the ${\rm HH_S}$ gauge is the most straightforward extension beyond the minimal gauge. Within our formalism, one notes that the gauge simply introduces an additional linear term $\sim S\sigma$ to the height function, beyond its leading term. However, the original formulation does not allow for a smooth transition between the MG ($S=0$) and the gauge ${\rm HH_S}$. Therefore, we proposed a modification in the ${\rm HH_S}$ to continuously recover the MG gauge in the limit $S\rightarrow0$, without changing the core properties of the ${\rm HH_S}$ gauge for $S\neq 0$ --- see eq.~\eqref{eq:mHHS}.

Despite the absence of studies of the hyperboloidal formalism in the frequency domain, refs.~\cite{Dolan:2009nk,Dolan:2010wr} introduced new regularisation factors for the equations. These factors were re-interpreted here as specific choices for the hyperboloidal coordinates in the spacetime picture. Apart from discussing such interpretation, this paper also reviews the constant mean curvature (CMC) and asymptotically constant mean curvature (ACMS) gauges, for which the height function depends on the angular coordinate $\theta$. A separation of the TE in the frequency domain is not available in such $\theta$-dependent gauges.
\medskip
s
By geometrically adapting the time coordinate to the black-hole horizon and the (infinitely) far wave zone, the hyperboloidal approach provides a robust formalism to black-hole perturbation theory. Specifically, the framework recast the quasi-normal problem in terms of the spectral problem of non-selfadjoint operators. Therefore, novel tools become available to develop the theory further. Among several possibilities, the spectral decomposition of the solutions to Teukolsky equation in terms of a quasi-normal mode (+ tail) expansion should be explored in future works. The semi-analytical algorithms from refs.~\cite{Ansorg2016,Macedo2018} applies directly to both the angular and radial equations in the frequency domain, while more rigours results on this topic can be explored along the lines of refs.~\cite{Gajic:2019oem,Gajic:2019qdd}. This work also lays the path for further applications of the hyperboloidal formalism in the Kerr spacetime, for instance in the context of the Lorenz gauge field equations~\cite{Dolan:2011dx,Dolan:2012jg,Dolan:2019hcw}. In a broader sense, the generic hyperboloidal framework is expected to contribute to studies on the EMRI problem via complementary approaches in the time and frequency domain.

\section*{Acknowledgments}
This work was supported by the European Research Council Grant No. ERC-2014-StG 639022-NewNGR ``New frontiers in numerical general relativity".

\appendix{
\section{Conformal Metric in hyperboloidal coordinates}\label{App:Metric}
Explicitly, the non-vanishing components of the conformal metric, its inverse and the determinant read
\bea
 \tilde{g}_{00} &=& -\sigma^2 F, \,\,   \tilde{g}_{01} = -(h_*+\beta), \,\, \tilde{g}_{02} = \sigma^2F h_{,\theta}, \,\, \tilde{g}_{03} = -\alpha \mu q \sin^2\theta. \nn \\
 \tilde{g}_{11}&=& h_{,\sigma}(h_*+2\beta), \,\, \tilde{g}_{12} = h_{,\theta}(h_*+\beta), \,\, \tilde{g}_{13} = \alpha \sin^2\theta \left(\beta +\mu  q h_{,\sigma} \right) \label{eq:ConfMetric}\\ 
 \tilde{g}_{22} &=& \tilde\Sigma - \sigma^2Fh_{,\theta}^2, \,\, \tilde{g}_{23} =  \alpha \mu q h_{,\theta} \sin^2\theta, \,\,
  \tilde{g}_{33} = \dfrac{\sin^2\theta}{\tilde\Sigma}\left( \tilde\Sigma_0^2 - \tilde\Delta \alpha^2\sigma^2\sin^2\theta\right); \nn
\eea
\bea
\tilde{g}^{00} &=& -\dfrac{h_{,\sigma}}{\beta^2\tilde\Sigma}\left( 2\beta\tilde\Sigma_0 - \sigma^2\tilde\Delta h_{,\sigma} \right) + \dfrac{\gamma}{\tilde\Sigma}, \,  \tilde{g}^{01} = \dfrac{\sigma^2\tilde\Delta h_{,\sigma}}{\beta^2\tilde\Sigma} - \dfrac{\tilde\Sigma_0}{\beta\tilde\Sigma},\,
 \tilde{g}^{02} = \dfrac{ h_{,\theta}}{\tilde\Sigma}, \label{eq:ConfInvMetric} \\
\tilde{g}^{03} &=& \dfrac{\alpha}{\tilde\Sigma}\left(1 - \dfrac{\sigma^2h_{,\sigma}}{\beta}\right), \, \tilde{g}^{11} = \dfrac{\sigma^2\tilde\Delta}{\beta^2 \tilde\Sigma}, \, 
\tilde{g}^{13} = -\dfrac{\alpha \sigma^2 }{\beta\tilde\Sigma}, \, \tilde{g}^{22} = \dfrac{1}{\tilde\Sigma}, \, \tilde{g}^{33} = \dfrac{1}{\tilde\Sigma\sin^{2}\theta}; \nn
\eea
\bea
\label{eq:ConfDet}
\tilde g &=& \det\tilde g_{ab} =  -\beta^2 \, \tilde\Sigma^2 \, \sin^2\theta.
\eea
For simplicity, we introduced the following quantities in the above expressions,
\beq
q = \dfrac{2\rho\sigma^3}{\tilde\Sigma}, \quad \gamma = h_{,\theta}^2 + \alpha^2\sin^2\theta, \quad h_* = -\sigma^2 Fh_{,\sigma}.
\eeq

\section{Length scales and dimensionless parameters}\label{App:LengthScale}
In this work, we employed the black-hole horizon $r_+$ as the typical length scale $\lambda$. We recall, however, that many works set $M=1$ in the numerical simulations, which is equivalent to the choice $\lambda = M$. Besides, Leaver~\cite{Leaver85} normalises according to $\lambda=2M$, whereas our previous works~\cite{Macedo2014,Ansorg2016,Macedo2018} had $\lambda=2r_+$. 

\begin{table}[h]
\caption{Dimensionless mass and spin parameters dependence on $\kappa$}
\begin{center}
\begin{tabular}{|c||c|c|}
\hline
$\quad \lambda \quad$ & $\quad \mu \quad$ & $\alpha$ \\
\hline
\hline
$M$ & $1$& $2\kappa/(1+\kappa^2)$ \\
\hline
$2M$ & $1/2$& $\kappa/(1+\kappa^2)$ \\
\hline
$r_+$ & $(1+\kappa^2)/2$& $\kappa$ \\
\hline
$2r_+$ & $(1+\kappa^2)/4$& $\kappa/2$ \\
\hline
\end{tabular}
\end{center}
\label{table:dimensonlessPar}
\end{table}

Exploiting the definition of the Kerr parameter $\kappa\in[0,1]$ introduced in eq.~\eqref{eq:KerrParm}, table~\ref{table:dimensonlessPar} brings the dependence on $\kappa$ of the dimensionless mass $\mu$ and spin $\alpha$ parameters according to each normalisation.

\section{Teukolsky equation in the minimal gauge}\label{App:TE_MG}
Here, we display the TE in the time and frequency domain for the two possible choices within the minimal gauge. We recall that the results are displayed in terms of the parameter $\kappa$ --- see eqs.~\eqref{eq:KerrParm} and \eqref{eq:kappa_a_over_M}. 
\subsection{Radial function fixing}
In the time domain, the TE \eqref{eq:Teukosly_Vm} in the MG$_{\rm R}$ gauge reads
\bea
\Bigg(4 \left(1+\kappa ^2\right) \left[1+\kappa ^2(1-   \sigma) \right] \left(1 +  \sigma[1+\kappa ^2] \right)- \kappa ^2 \left(1-x^2\right) \Bigg)V_m{}_{,\tau \tau} \nn \\
-2\Bigg(  1+\kappa ^2 \sigma ^2 - 2 \sigma^2 \left(1+\kappa ^2\right)  \left[1 + \kappa ^2 (1-  \sigma) \right] \Bigg) V_m{}_{,\tau \sigma} -(1-x^2) V_m{}_{,xx}\nn \\
-\sigma ^2(1-\sigma )  \left(1-\kappa ^2 \sigma \right) V_m{}_{,\sigma \sigma}
+2\Bigg[2\sigma \left(1+\kappa ^2\right)    \left[1+\kappa ^2(1-2  \sigma) \right]  \nn \\
- \kappa ^2 \sigma  \left[1-\sigma(1+\kappa ^2  ) \right]  -p \left[\left(1+\kappa ^2\right)   \left[1- \sigma(1 +\kappa ^2) \right]- i \kappa  x\right]   \nn \\
 + i \kappa  m \left[1 + 2 \sigma   \left(1+\kappa ^2\right)  \right]  \Bigg] V_m{}_{,\tau} + \Bigg(2 x - \delta_1 (1-x)+\delta_2(1+x)\Bigg)V_m{}_{,x} \label{eq:TE_RadFix} \\
-\sigma  \Bigg(2 (1+p)-\sigma  \left[2 i \kappa  m+\left(1+\kappa
   ^2\right) (3+p)-4 \kappa ^2 \sigma\right]\Bigg)  V_{m}{}_{,\sigma} +\Bigg[ \left(1+\kappa^2\right) p \sigma\nn \\
  \sigma  \left[1+\kappa ^2(1-2 \sigma) \right]+2 i \kappa  m \sigma
   -\left(p-\frac{\delta_1+\delta_2}{2}\right)
   \left(1+p+\frac{\delta_1+\delta_2}{2}\right)      \Bigg]V_m =0\nn
\eea
}
whereas the radial operator~\eqref{eq:QNM_Hyp} in the frequency domain is
\bea
{\boldsymbol A} = \sigma ^2(1-\sigma )  \left(1-\kappa ^2 \sigma \right)\dfrac{\d^2}{\d \sigma^2} 
+ \Bigg[ 2 (p \sigma +s+\sigma )
+\sigma ^2 \Bigg(-2 i \kappa  m-\left(1+\kappa ^2\right) p \nn \\
-3 \left[1+\kappa ^2(1+2  s) \right]+4 \left[\kappa ^2 \sigma -\kappa ^4 s (1-\sigma)-s \left(1-\kappa ^2 \sigma \right)\right]\Bigg) \Bigg] \dfrac{\d}{\d \sigma} \nn \\
-A_{\ell m}-\sigma  \left[1+2s(1 + \kappa ^2)\right] \left[2 i \kappa  m + \left(1+\kappa ^2\right) p\right]-2 s \left[i \kappa  m - p \left(1+\kappa ^2\right) \right] \nn \\
+\kappa ^2 s^2-4 \left(\kappa ^2+1\right)^2 s^2+2 \kappa ^2 \sigma ^2 \left(\kappa ^2s+s+1\right) \left[1+2 s \left(1+\kappa ^2\right) \right] \nn \\
-\sigma  \left[1+2s \left(1+\kappa^2\right) \right] \left[(1+\kappa ^2)(1+2s)+ 2s\kappa ^4\right] 
\eea

\subsection{Cauchy horizon fixing}
Finally, we present the Teukolsky equation in the MG$_{\rm C}$. In the time domain, it reads
\bea
\Bigg[ 4 \left(1+\kappa ^2\right) \left[1+\frac{\left(1+\kappa ^2\right) \sigma }{1-\kappa^2}\right]-\kappa ^2 \left(1-x^2\right) \Bigg] V_m{}_{,\tau \tau} - \sigma ^2(1-\sigma) \, V_m{}_{,\sigma \sigma}\nn \\
-2 \left[ (1-\sigma ) \left[ 1+\sigma -\kappa^2(1- \sigma)\right]-\frac{\left(1+\kappa ^2\right) \sigma ^2}{1-\kappa ^2}\right] V_m{}_{,\tau \sigma} - (1-x^2)V_m{}_{,xx}\nn \\
-\sigma \Bigg[(1+p) (2-\sigma)  -2 \sigma\left(1+\frac{ i \kappa  m }{1-\kappa ^2}\right)\Bigg]V_m{}_{,\sigma}   + \Bigg[2 x - \delta_1 (1-x) \nn \\ 
+\delta_2(1+x)\Bigg]V_m{}_{,x} + \Bigg[2 \kappa  (i m-\kappa) +2 p \left[i \kappa  x -\left(1+\kappa ^2\right) (1-\sigma )\right] \nn \\ 
+ \frac{2 \sigma \left(1+\kappa ^2\right)   \left(2 + 2 i \kappa  m - \kappa ^2\right)}{1-\kappa ^2}\Bigg] V_m{}_{,\tau} +\Bigg[\sigma\left (1+p + \frac{2 i \kappa  m  }{1-\kappa ^2}\right) \nn \\
-\left(p-\frac{\delta_1+\delta_2}{2}\right)  \left(1+p+\frac{\delta_1+\delta_2}{2}\right) \Bigg] V_m =0. \label{TE:CauchyFix} 
\eea
whereas the radial operator in the frequency domain is
\bea
 \sigma ^2(1-\sigma) \,  \dfrac{\d^2}{\d \sigma^2}
 + \Bigg[ (2-\sigma) \sigma  \left(1+p+2 \kappa ^2 s\right)-2 \sigma ^2 + 2 \left(1-\kappa ^2\right) s \nn \\
 -\frac{2 \sigma ^2 (2 s+i \kappa  m)}{1-\kappa ^2} \Bigg] \dfrac{\d}{\d \sigma}   -A_{\ell m}+s \left[2 p+ \kappa^2s - 2 i m \kappa \right]-(1+p) \left(\sigma-2 \kappa ^2 s \right) \nn \\
 -2 s \left(1+\kappa ^2\right) \left[ (1+p) \sigma +2 s\right] - \frac{2 \sigma  \left[1+2 \left(1+\kappa ^2\right)  s\right] \left[\left(1+\kappa ^2\right) s+i \kappa  m\right]}{1-\kappa ^2}. \nn
\eea

\section*{Bibliography}
\bibliographystyle{unsrt.bst}
\bibliography{bibitems}

\end{document}